\documentclass{article}

\usepackage{arxiv}

\usepackage[utf8]{inputenc} 
\usepackage[T1]{fontenc}    
\usepackage{hyperref}       
\usepackage{url}            
\usepackage{booktabs}       
\usepackage{amsfonts}       
\usepackage{nicefrac}       
\usepackage{microtype}      
\usepackage{graphicx}
\usepackage{natbib}
\usepackage{doi}
\usepackage{xcolor}
\usepackage{enumerate}
\usepackage{amsmath}
\usepackage{braket}
\usepackage{subcaption}

\newcommand{\mc}{\mathcal}
\newcommand{\bs}{\boldsymbol}
\newcommand{\R}{\mathbb{R}}
\newcommand{\C}{\mathbb{C}}

\newtheorem{remark}{Remark}[section]

\DeclareMathOperator{\supp}{supp}
\DeclareMathOperator{\tr}{Tr}

\title{Piecewise Polynomial Tensor Network Quantum Feature Encoding}

\author{ \href{https://orcid.org/0000-0003-1664-7098}{\includegraphics[scale=0.06]{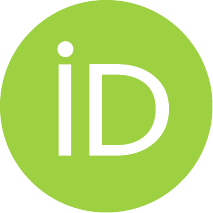}\hspace{1mm}Mazen Ali} \\
        Formerly with the Department of Flow and Material Simulation\\
	Fraunhofer ITWM\\
	Kaiserslautern, Germany 67663 \\
	\texttt{mazen.ali90@gmail.com} \\
	\And
	\href{https://orcid.org/0000-0001-8869-2387}{\includegraphics[scale=0.06]{orcid.pdf}\hspace{1mm}Matthias Kabel} \\
	Department of Flow and Material Simulation\\
	Fraunhofer ITWM\\
	Kaiserslautern, Germany 67663 \\
	\texttt{matthias.kabel@itwm.fraunhofer.de} \\
}

\renewcommand{\headeright}{}
\renewcommand{\undertitle}{}
\renewcommand{\shorttitle}{PPTNQFE}

\hypersetup{
pdftitle={PPTNQF Encoding},
pdfsubject={quanth-ph, cs.LG},
pdfauthor={Mazen Ali, Matthias Kabel},
pdfkeywords={Quantum Machine Learning},
}

\begin{document}
\maketitle

\begin{abstract}
This work introduces a novel method for embedding continuous variables into quantum circuits via piecewise polynomial features, utilizing low-rank tensor networks. Our approach, termed Piecewise Polynomial Tensor Network Quantum Feature Encoding (PPTNQFE), aims to broaden the applicability of quantum algorithms by incorporating spatially localized representations suited for numerical applications like partial differential equations and function regression. We demonstrate the potential of PPTNQFE through efficient point evaluations of solutions of discretized differential equations and in modeling functions with localized features such as jump discontinuities. While promising, challenges such as unexplored noise impact and design of trainable circuits remain.
\end{abstract}

\keywords{Quantum Machine Learning \and Feature Encoding \and Tensor Networks \and Piecewise Polynomial \and Splines \and Differential Equations \and Finite Element Methods}

\section{Introduction}
In the quest for viable quantum computing, Quantum Machine Learning (QML) has emerged as a rapidly expanding domain of research.
The aspiration is that a model could discern patterns more effectively by navigating the quantum state space, as opposed to relying solely on classical states.
It appears reasonable to assert that this methodology is advantageous for data emanating from inherently quantum processes, and there is now substantial evidence supporting this claim \cite{Preskill21, Preskill22}.
Nonetheless, given that the majority of data-driven applications do not operate within the quantum realm, the pertinent question remains: does QML hold utility for classical data and problems?

\subsection{Quantum Advantage in Quantum Machine Learning}
A central theme in Quantum Machine Learning (QML), and in quantum computing at large, is the pursuit of the elusive ``quantum advantage'' (QA).
This term is broadly defined as the superior performance of a quantum model over its best classical counterpart.
Despite the numerous claims of achieving QA, practical applications remain conspicuously absent \cite{Preskill23, Abbas2023, Montanaro16, Ali23}.

Theoretically, there exist rigorous demonstrations of QA in specific learning contexts \cite{Liu21, Muser23, Yamasaki23}.
However, these proofs are often applicable to contrived problems lacking real-world relevance.
On the practical front, the capabilities of contemporary quantum devices are limited by their small scale and significant noise levels, rendering them inadequate for large-scale demonstrations.
Efficiently scaling quantum model training remains debatable \cite{Abbas2023}, yet our focus here is more on generalization, especially through the lens of feature embeddings.

The reevaluation of QA as the primary goal for QML has been insightfully suggested by \cite{Schuld22}.
A more constructive approach may involve exploring the unique characteristics of quantum models and identifying effective foundational elements. Despite this, there remains a significant focus within the literature on developing intricate hybrid models.
However, even among these advanced models, the underlying building block often remains the relatively straightforward concept of rotation-based
parametrized quantum circuits (PQCs), see Figure \ref{fig:embeddings}.

The rationale for employing PQCs often hinges on the large size of the quantum state space.
Nonetheless, these circuits often produce Fourier series models, which may not fully take advantage of the exponentially large Hilbert space and can be emulated classically \cite{Schuld21, Kordzanganeh23, Shin23},
or they may be overly expressive, hindering effective training and generalization \cite{Kubler21, Huang21, Jerbi23, Cerezo23}.
Furthermore, the impact of statistical and hardware-induced errors is frequently overlooked, potentially undermining any claims of enhanced expressiveness or precision.

\subsection{Inductive Bias and Feature Encoding}
The efficacy of a model extends beyond mere expressiveness; it depends on the synergy between \emph{sufficient} expressiveness and
the appropriate, problem-specific \emph{inductive bias}.
Recent innovative studies have explored the potential to encode specific biases more efficiently using quantum models
\cite{Bowles23, Gili23, Nguyen22, Schatzki24}.

A pivotal aspect common to both classical and quantum models is feature encoding, the process by which input features are integrated into the model's latent spaces, enabling the training algorithm to detect patterns. Several promising methods have emerged to embed biases through feature encoding in QML
\cite{Peters23, QCNN, Meyer23, Umeano23, Umeano24}.
Yet, the prevalent practice in QML models, especially those dealing with classical data, remains the use of rotation-based encodings.
This approach inherently predisposes the model to identify wave-like, global,
periodic patterns, which, while beneficial for certain tasks, constitutes a constraint for many others.

\subsection{This Work}
To broaden the spectrum of quantum feature embeddings (QFEs), this work introduces an approach to efficiently integrate piecewise polynomial features into quantum models. Piecewise polynomial approximations are widely utilized in various numerical applications, including interpolation and solving partial differential equations (PDEs), as well as in classical deep learning frameworks like networks employing rectified linear units (ReLUs). Spatially localized representations, such as those provided by piecewise polynomials, are often more suitable for certain applications compared to global trigonometric polynomials
\cite{Schuld21} or Chebyshev polynomials \cite{Kyriienko21, Williams23}.

Our approach leverages an intermediate tensor network (TN) representation, which makes it adaptable to any function that can be represented in a low-rank format
\cite{Oseledets13, AliNouy23}.
We denote these embeddings as \underline{\textbf{P}}iecewise \underline{\textbf{P}}olynomial
\underline{\textbf{T}}ensor \underline{\textbf{N}}etwork \underline{\textbf{Q}}uantum \underline{\textbf{F}}eature \underline{\textbf{E}}ncodings
(PPTNQFE). It is crucial to note that these embeddings are not inherently superior or inferior to others and do not confer any QA. Through toy examples, we elucidate the scenarios where PPTNQFEs are advantageous and when they are not.

The structure of this paper is as follows.
Section \ref{sec:pp} reviews piecewise polynomials and the properties required for embedding.
Section \ref{sec:tn} discusses TNs and functional TN representations.
Section \ref{sec:meth} details the methodology for expressing piecewise polynomial bases as low-rank TNs and their subsequent conversion into quantum circuits.
Section \ref{sec:app} presents three important applications: point evaluations of PDE solutions, function learning (regression) and classification with quantum
kernels.
The paper concludes in Section \ref{sec:concl}.
All computations with quantum algorithms were performed
using a state vector simulator.

\section{Piecewise Polynomials}\label{sec:pp}
Within a bounded domain $\Omega\subset\R^d$ and a partition $\mc T_{\tilde{N}}:=\{T_0,\ldots, T_{\tilde{N}-1}\}$  that collectively covers
$\overline{\Omega}=\bigcup_i T_i$,
functions $S_N:=\{\varphi_0,\ldots,\varphi_{N-1}\}$ are said to be \emph{piecewise polynomial} if the restriction $\varphi_k|_{T_i}$ is a polynomial for every $k$ and $i$.
This framework is typical in finite element methods (FEM) \cite{Braess07}.
Similarly, in learning, (leaky) ReLU networks yield piecewise polynomial functions \cite{Yarostsky17}.
For numerical stability and network trainability, these functions are usually required to satisfy
global regularity, $\varphi_k\in C^m(\Omega)$, with $m\geq 0$.
This in particular excludes discontinuous TN polynomial features as in \cite{AliNouy23}.
While TN approximations might offer implicit regularization through low ranks (see examples in \cite{Nouy23}),
embedding discontinuous features into quantum circuits complicates training.

Two essential properties are stipulated for this set of functions:
\begin{enumerate}[(I)]
    \item\label{req:fixed} For any given partition $\mc T_{\tilde{N}}$, each element $T_i$ is predetermined, i.e., parameter-independent, and its coordinates can be computed in $\mc O(1)$.
    This excludes replicating the ReLU approximations as in \cite{Yarostsky17}.
    \item\label{req:local} The functions in $S_N$ have local support:
    \begin{align*}
        \#\left\{k=0,\ldots,N-1:|\supp \varphi_k\cap\supp\varphi_i|>0\right\}=\mc O(1),
    \end{align*}
    for any $i$. In other words, for any given $x\in\overline{\Omega}$, there is only a constant number of $\varphi_k$'s such that $\varphi_k(x)\neq 0$.
\end{enumerate}
The locality property \eqref{req:local} leads to sparse embedding vectors.
Instead of sparsity, we could require any set of functions that are representable with low-rank TNs such as, e.g., ``saw-tooth'' functions, see
\cite{AliNouy23}.

Additionally, though not mandatory, it is beneficial for $S_N$ to form a partition of unity:
\begin{align}
    \sum_k\varphi_k(x)=1,\quad\forall x\in\overline{\Omega}\tag{III}\label{req:unity}.
\end{align}

\begin{figure}[ht!]
  \centering
  \includegraphics[height=7.5cm, keepaspectratio]{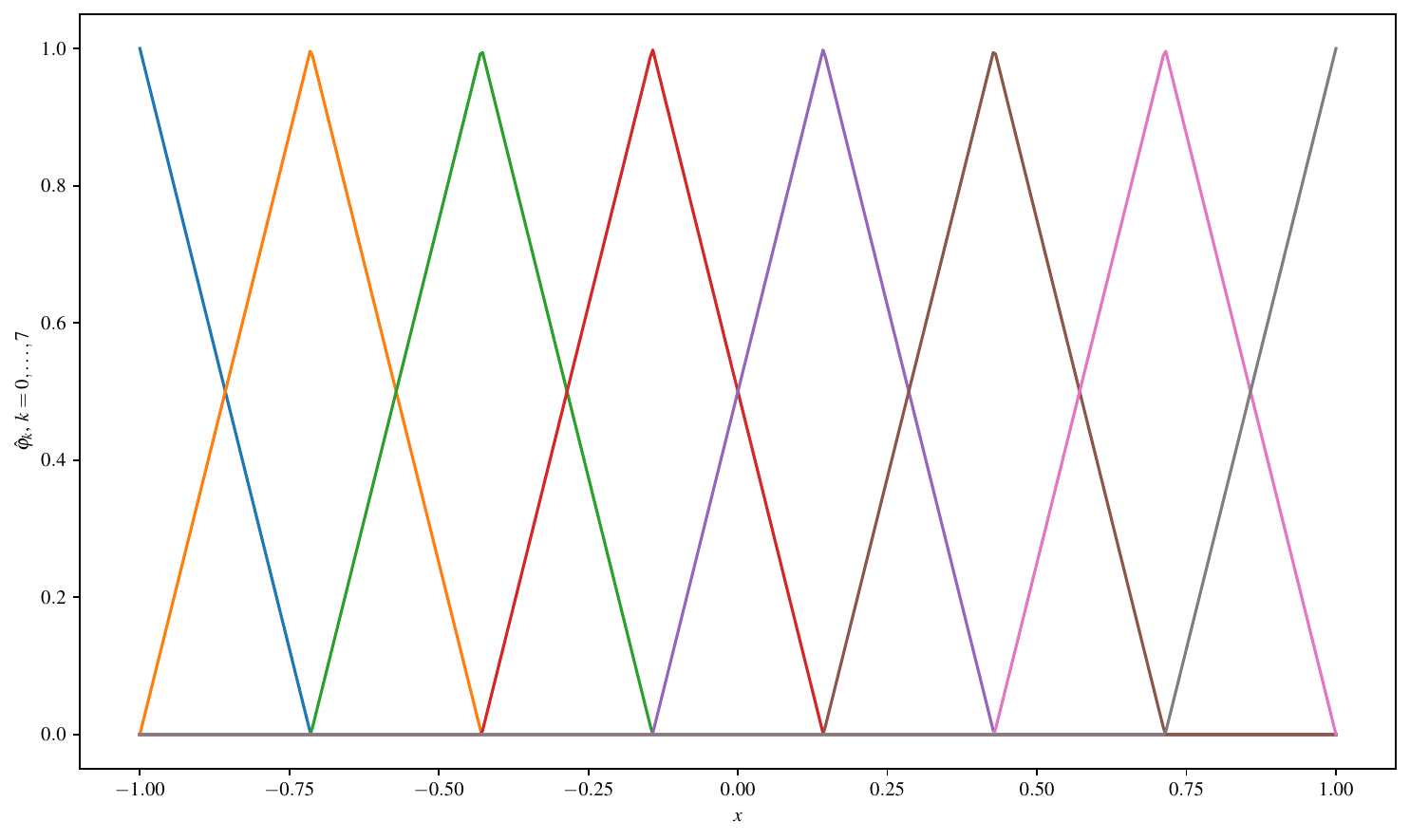}
  \caption{1D piecewise linear hat basis for a grid of 8 uniformly spaced points, including the boundary $x\in\{\pm 1\}$.}
  \label{fig:hat_basis}
\end{figure}

\begin{figure}[ht!]
  \centering
  \includegraphics[height=7.5cm, keepaspectratio]{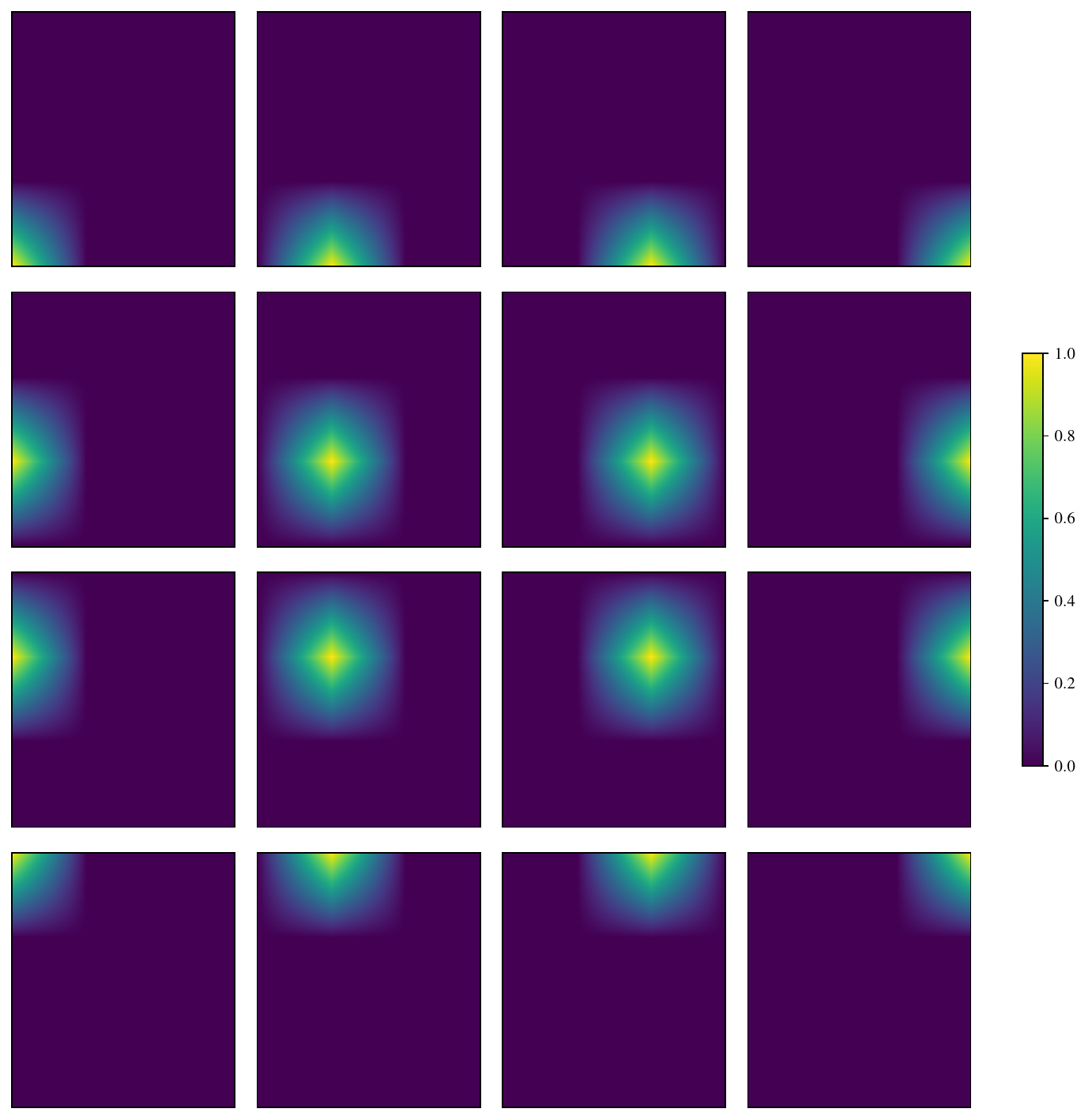}
  \caption{2D bilinear basis functions formed as tensor products of 4 linear basis
  functions in each coordinate direction, i.e., 16 functions in total.}
  \label{fig:hat_basis2D}
\end{figure}

In this work, we will explicitly derive and implement an embedding for a 1D basis on the interval $\overline{\Omega} = [-1, 1]$.
The basis consists of piecewise linear hat functions
\begin{align*}
    \hat{\varphi}_k(x)=
    \begin{cases}
        \frac{x-x_{k-1}}{h},&\quad\text{for }x_{k-1}\leq x < x_k,\\
        \frac{x_{k+1}-x}{h},&\quad\text{for }x_{k}\leq x < x_{k+1},\\
        0,&\quad\text{otherwise},
    \end{cases}
\end{align*}
for uniformly spaced $x_k$, where $h:=1/(N-1)$.
The placing of the nodes $x_k$ will depend on the application:
\begin{itemize}
    \item In Section \ref{sec:app:de}, $h=1/(2^n-1)$, $x_0=-1+h$, $x_{2^n-1}=1-h$, where $n$ is the number of qubits.
    \item In Section \ref{sec:app:reg}, $x_0=-1$, $x_{2^n-1}=1$.
\end{itemize}
A plot of the hat functions is presented in Figure \ref{fig:hat_basis}.
These satisfy conditions \eqref{req:fixed}, \eqref{req:local} and \eqref{req:unity}.
Moreover, by taking tensor products of the proposed feature embeddings, we can trivially extend them to any spatial dimension. See
Figure \ref{fig:hat_basis2D} for a two-dimensional example.

\section{Tensor Networks}\label{sec:tn}
\begin{figure}[htbp]
  \centering
  \begin{subfigure}[b]{0.49\textwidth}
    \includegraphics[width=\textwidth, trim={0 3cm 1cm 3cm}, keepaspectratio]{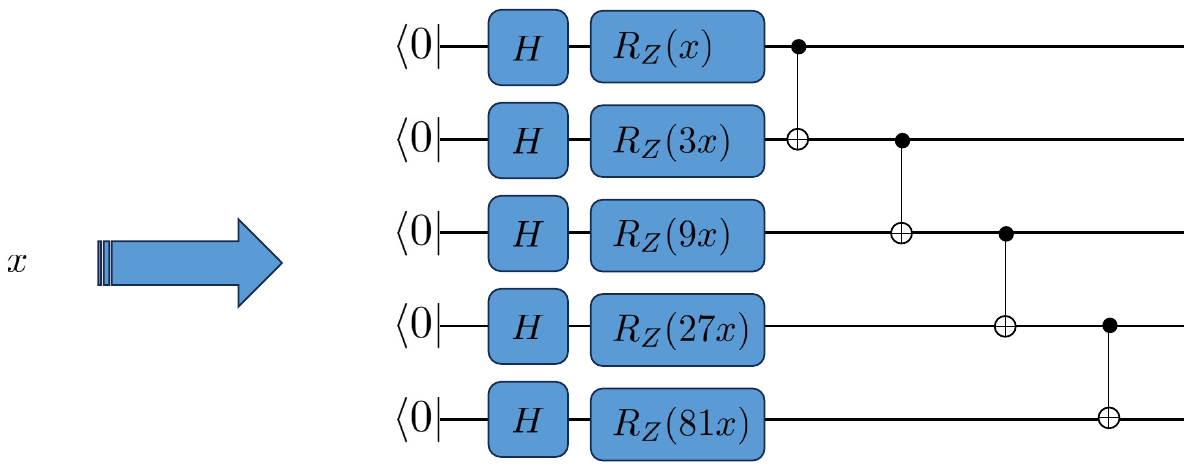}
  \end{subfigure}
  \begin{subfigure}[b]{0.49\textwidth}
    \includegraphics[width=\textwidth, trim={1cm 3cm 0 3cm}, keepaspectratio]{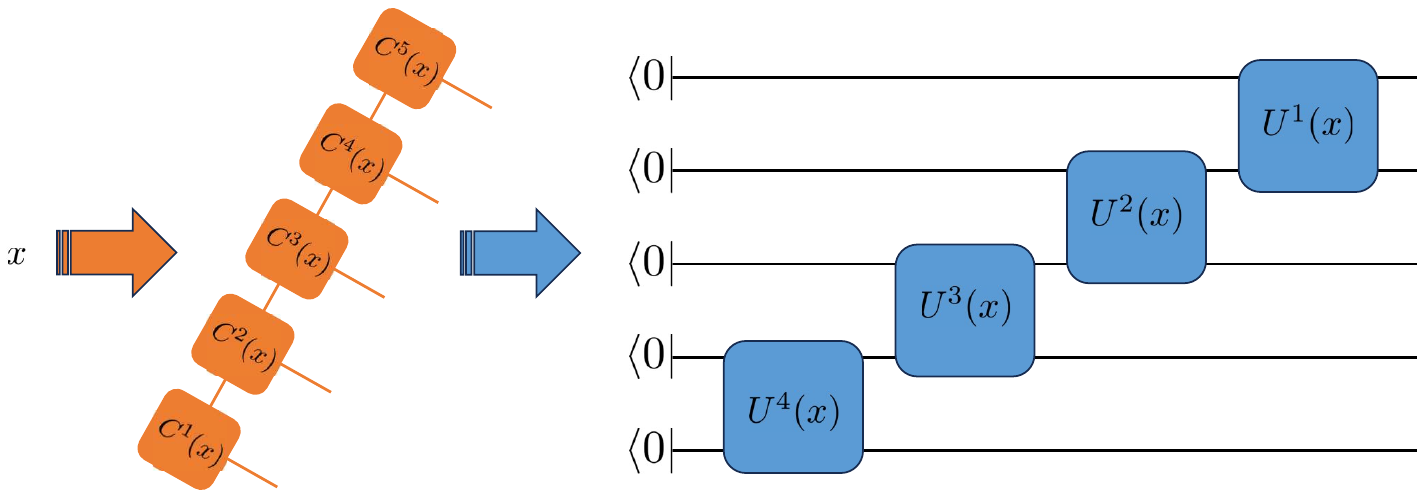}
  \end{subfigure}
  \begin{subfigure}[b]{0.49\textwidth}
    \includegraphics[width=\textwidth, trim={1cm 3cm 0 3cm}, keepaspectratio]{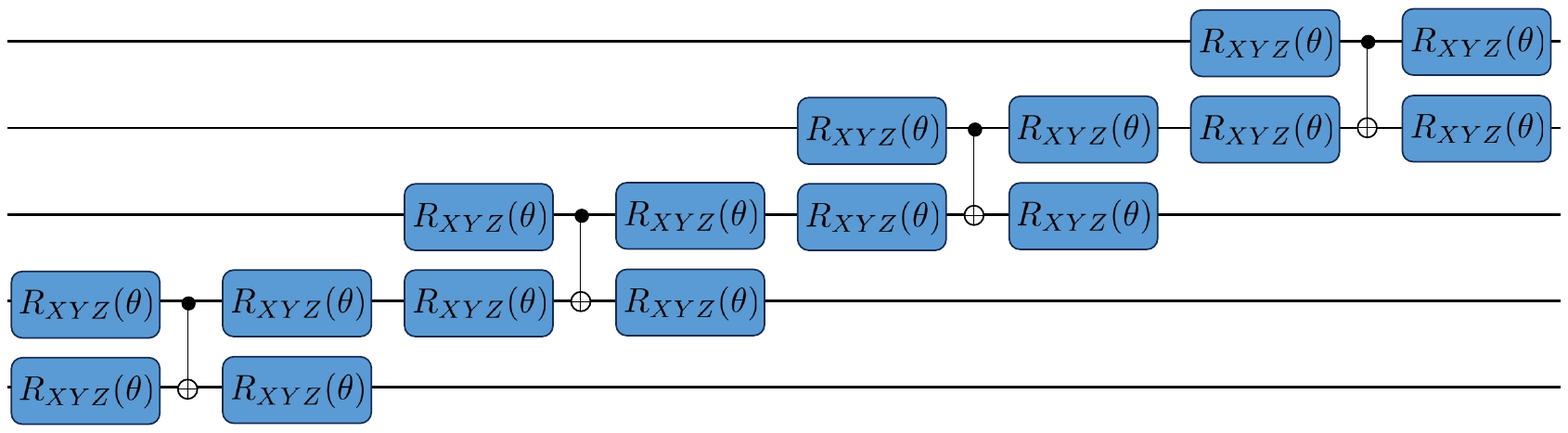}
  \end{subfigure}
  \caption{On the top left, an example of a rotation-based exponential encoding.
  These embeddings can be repeated, or applied to multidimensional features $x$ with multi-qubit rotations depending on products of features $R_{ZZ}(x_ix_j)$, see also \cite{Havlicek19}.
  On the top right, an encoding of the features into a matrix product state (MPS) with maximum rank 2 and converting the latter to a quantum circuit.
  On the bottom, an example of a variational MPS layer that we will require in Section
  \ref{sec:app:reg}.}
  \label{fig:embeddings}
\end{figure}

Tensor networks are an implicit, parsimonious representation of high-order tensors. They are most commonly known for their effectiveness in modeling
high-dimensional quantum systems with low entanglement. For details on TN formats and diagrams,
see \cite{Biamonte17}.
In this work, we will be using the matrix product state (MPS) format
\cite{Schollwock11}, also known under the name tensor train (TT) format \cite{TT},
see Figure \ref{fig:embeddings}.
A state $\ket{\psi}$ over $n$ sites is called an MPS
(with open boundary conditions) if it can be expressed as
\begin{align*}
    \ket{\psi}=\sum_{s_0,\ldots, s_{n-1}}C^0_{s_0}\cdots C^{n-1}_{s_{n-1}}
    \ket{s_0\ldots s_{n-1}}.
\end{align*}

A crucial feature of TNs is that they can be converted into quantum circuits.
This was done for MPS in \cite{Ran20}.
For this work, the implementation was based on the details specified in \cite{Lubasch20}, more details can be found in Section \ref{sec:data}.
The width of the multi-qubit gates depends logarithmically on the maximum MPS rank. For the MPS embedding presented in this work,
the maximum rank is 2 and thus we only require 2-qubit gates for the conversion in Figure \ref{fig:embeddings}.

For loading sparse vectors onto a quantum computer, we do not need TNs and could, in principle, use other methods as in \cite{Gleinig21}.
However, TNs offer advantages for function representations and approximations that could potentially allow the method to be generalized
to other sets of functions. There is an extensive theory behind TN properties for function approximation (see, e.g., \cite{Oseledets13, AliNouy23, Kazeev17}),
numerical methods (see, e.g., \cite{Kazeev17, Bachmayr20, Haberstich23, Nouy23, Ritter24}), and applications to quantum computing
(see, e.g., \cite{Holmes20, Melnikov23, Gourianov22, Miyamoto23, Iaconis24}).
Moreover, as mentioned in Section \ref{sec:pp}, we could also load nonlocal function sets such as saw-tooth functions.

\section{Methodology}\label{sec:meth}
Prior to transferring the basis functions from Section \ref{sec:pp} to a quantum computer,
a transformation $\varphi_k=g\circ\hat{\varphi}_k$ might be necessary. For instance, in Section \ref{sec:app:reg}, we apply
the transformation $\varphi_k=\sqrt{\hat{\varphi}_k}$.
We only require this transformation to preserve the locality property \eqref{req:local}.
Additionally, normalization might require monitoring the norm
\begin{align*}
    n_\Phi(x):=\left(\sum_k\varphi_k^2(x)\right)^{1/2}.
\end{align*}
Given the property \eqref{req:local}, $n_\Phi(x)$ can be computed in $\mc O(1)$ for any $x\in\overline{\Omega}$.
For the case $\varphi_k=\sqrt{\hat{\varphi}_k}$, the property \eqref{req:unity} ensures $n_\Phi(x)=1$,
facilitating automatic normalization of the basis vector.

We have the basis vector
\begin{align*}
    \Phi(x):=
    \begin{bmatrix}
        \varphi_0(x) &\\
        \vdots &\\
        \varphi_{2^n-1}(x)&
    \end{bmatrix},
\end{align*}
and we want to prepare the corresponding normalized quantum state,
\begin{align*}
    \ket{\Phi(x)}:=\frac{1}{n_\Phi(x)}\sum_{k=0}^{2^n-1}\varphi_k(x)\ket{k},
\end{align*}
where $\ket{k}$ are the computational basis states.
The basis vector $\Phi(x)$ is initially represented as an MPS.
For $n$ qubits, we write an integer index in binary form $k=i_0i_1\cdots i_{n-1}$,
where we use the Big Endian convention (the most significant bit is $i_0$).
Then, we seek a representation
\begin{align*}
    \Phi(x)=\left(\varphi_{i_0\cdots i_{n-1}}(x)\right)_{i_0=0,\ldots,i_{n-1}=0}^1=
    C^0(x)[\cdot,i_0,\cdot]\cdots C^{n-1}(x)[\cdot,i_{n-1},\cdot],
\end{align*}
where the order-3 \emph{cores} have dimensions $C^j(x)\in\C^{r_j\times 2\times r_{j+1}}$ and open boundary conditions $r_0=r_n=1$.

In order for the complexity of the embedding circuits to remain small, it is important the ranks $r_j$ remain small as well.
There are two properties crucial for this: the number of non-zero entries in $\Phi(x)$ and their relative position with respect to each other,
i.e., the choice of ordering for the basis functions with respect to $k=i_0\cdots i_{n-1}$.
Alternatively, as mentioned in Section \ref{sec:pp}, we could use non-local bases that have a low-rank representation.
We demonstrate how to represent the 1D hat basis from Section \ref{sec:pp} as a TN.

For any $x\in[-1,1]$, there are at most two adjacent non-zero values,
\begin{align*}
    \Phi(x):=
    \begin{bmatrix}
        0 &\\
        \vdots &\\
        \varphi_k(x)=:a &\\
        \varphi_{k+1}(x)=:b &\\
        \vdots &\\
        0&
    \end{bmatrix},
\end{align*}
The MPS representation varies depending on whether the index 
$k$ is even or odd, with a detailed explanation provided for the latter.
Let
\begin{align*}
    k&=i_0\cdots i_{j^*-1}0_{j^*}1_{j^*+1}\cdots1_{n-2}1_{n-1},\\
    k+1&=i_0\cdots i_{j^*-1}1_{j^*}0_{j^*+1}\cdots0_{n-2}0_{n-1},
\end{align*}
where $j^*$ is the position of the least significant $0$-bit in $k$.
The nonzero entries of the MPS cores are given as
\begin{align*}
    C^j(x)&\in\R^{1\times 2\times 1},\\
    C^j(x)[0, i_j, 0]&=1,
\end{align*}
for $j=0,\ldots,j^*-1$,
\begin{align*}
    C^j(x)&\in\R^{1\times 2\times 2},\\
    C^j(x)[0, 0, 1]&=1,\\
    C^j(x)[0, 1, 0]&=1,\\
\end{align*}
for $j=j^*$,
\begin{align*}
    C^j(x)&\in\R^{2\times 2\times 2},\\
    C^j(x)[0, 0, 0]&=1,\\
    C^j(x)[1, 1, 1]&=1,
\end{align*}
for $j=j^*+1,\ldots,n-2$, and
\begin{align*}
    C^{n-1}(x)&\in\R^{2\times 2\times 1},\\
    C^{n-1}(x)[0, 0, 0]&=b,\\
    C^{n-1}(x)[1, 1, 0]&=a.
\end{align*}
The case of even $k$ is similar.
Since the maximum rank is $2$, converting to a quantum circuit (see Figure \ref{fig:embeddings})
we require only $2$-qubit gates.

\begin{remark}
    We stress the difference from the work in \cite{Holmes20, Melnikov23, Miyamoto23, Iaconis24}: here we load the set of functions $\{\varphi_k(x)\}_{k=0}^{N-1}$, evaluated at a point $x\in\overline{\Omega}$, instead of loading a discrete representation of a single function $\{f_k=f(x_k)\}_{k=0}^{N-1}$.
\end{remark}

Finally, having embedded 1D features, we can easily extend this approach to any number of features by taking tensor products of the MPSs. For instance, given three features \(x\), \(y\), and \(z\) with corresponding binary indices \(i_k\), \(j_k\), and \(m_k\), one possible construction is the \emph{canonical} ordering:
\begin{align*}
    \Phi(x, y, z)=
    &A^0(x)[\cdot,i_0,\cdot]\cdots A^{n-1}(x)[\cdot,i_{n-1},\cdot]\cdot\\
    &B^0(y)[\cdot,j_0,\cdot]\cdots B^{n-1}(y)[\cdot,j_{n-1},\cdot]\cdot\\
    &C^0(z)[\cdot,m_0,\cdot]\cdots C^{n-1}(z)[\cdot,m_{n-1},\cdot].
\end{align*}
Another commonly used alternative is the \emph{transposed}
ordering
\begin{align*}
    \Phi(x, y, z)=
    &A^0(x)[\cdot,i_0,\cdot]\cdot B^0(y)[\cdot,j_0,\cdot]\cdot
    C^0(z)[\cdot,m_0,\cdot]\cdots\\
    &A^{n-1}(x)[\cdot,i_{n-1},\cdot]\cdot B^{n-1}(y)[\cdot,j_{n-1},\cdot]\cdot
    C^{n-1}(z)[\cdot,m_{n-1},\cdot].
\end{align*}
For further details on the effects of different orderings,
see \cite{Kazeev17, markeeva2021, AliNouy23}.
All orderings yield a feature vector \(\Phi(x, y, z)\) with the same values but arranged in different sequences.

\begin{remark}
    We emphasize that constructing the MPS for this feature encoding is computationally efficient, with a complexity of $\mc{O}(r)$. Specifically, in the example discussed in this paper, where $r = 2$, the construction is particularly inexpensive.
\end{remark}

\section{Applications}\label{sec:app}
In this section, we dive into three illustrative applications of the PPTNQFE method that we have introduced.
The first application focuses on the efficient point evaluation of solutions to differential equations.
In this scenario, the discretized and normalized solution is encoded in a quantum state, with the basis coefficients represented within the quantum amplitudes.
The second application we explore is regression, specifically the supervised learning of functions depending on continuous variables from given data samples.
Here, PPTNQFE allows, e.g., the approximation of jump functions in a particularly simple manner.
Finally, the third application is classification with quantum kernels. We demonstrate on synthetic datasets that, in some cases, PPTNQFE achieves similar results to rotation-based encoding. In other cases, PPTNQFE can avoid misclassification due to
the periodic nature of rotation-based encodings.

\subsection{Differential Equations}\label{sec:app:de}
The prevalent methodologies for solving PDEs on quantum computers involve mapping the problem's degrees of freedom into the amplitudes of a quantum state. This strategy has been applied to various types of PDEs, including linear and nonlinear, as well as stationary and time-dependent cases. In both quantum and classical digital computing, the initial step involves discretizing the infinite-dimensional PDE problem into a finite-dimensional system. Within the framework of FEM, a widely used discretization technique, the approximate solution is represented as
\begin{align*}
    u\approx\sum_k\bs u_k\varphi_k,
\end{align*}
where $\varphi_k$ are piecewise polynomials as described in Section \ref{sec:pp}, and $\bs u=(\bs u_k)_k$ is the solution vector containing the
degrees of freedom.

The next step involves solving the potentially nonlinear system 
\begin{align}\label{eq:pde_discrete}
    \bs A(\bs u)=\bs f,
\end{align}
where $\bs A$ and $\bs f$ represent the differential operator, the right-hand side (RHS), and possibly the boundary conditions, all formulated in the basis 
$S_N$.
On a quantum computer, solving \eqref{eq:pde_discrete} can be approached using algorithms such as the HHL algorithm
\cite{HHL1, HHL2} or variational methods \cite{Bravo23, Lubasch20}.
The outcome is a quantum state 
\begin{align*}
    \ket{\bs u}:=\frac{\bs u}{\|\bs u\|}=\frac{1}{\|\bs u\|}\sum_k\bs u_k\ket{k},
\end{align*}
or an approximation thereof.

In quantum computing, unlike classical computing, the coefficients of $\bs u$ are encoded in the state's amplitudes and are not directly accessible.
Typically, one can only measure observables related to $\ket{\bs u}$. However, for point evaluations of the solution $u(x)$,
one can use the relationship
\begin{align*}
    u(x)=\sum_k\bs u_k\varphi_k(x)=\|\bs u\|n_\Phi(x)\Re\braket{\bs u|\Phi(x)}.
\end{align*}
where $n_\Phi(x)$ is either $1$ or can be efficiently calculated in $\mc O(1)$.
Utilizing preparation procedures for $\ket{\Phi(x)}$ and $\ket{\bs u}$,  and employing techniques like the Hadamard test,
one can efficiently estimate
$\Re\braket{\bs u|\Phi(x)}$ and thus $u(x)$.

\begin{figure}[ht!]
  \centering
  \includegraphics[width=0.9\textwidth, keepaspectratio]{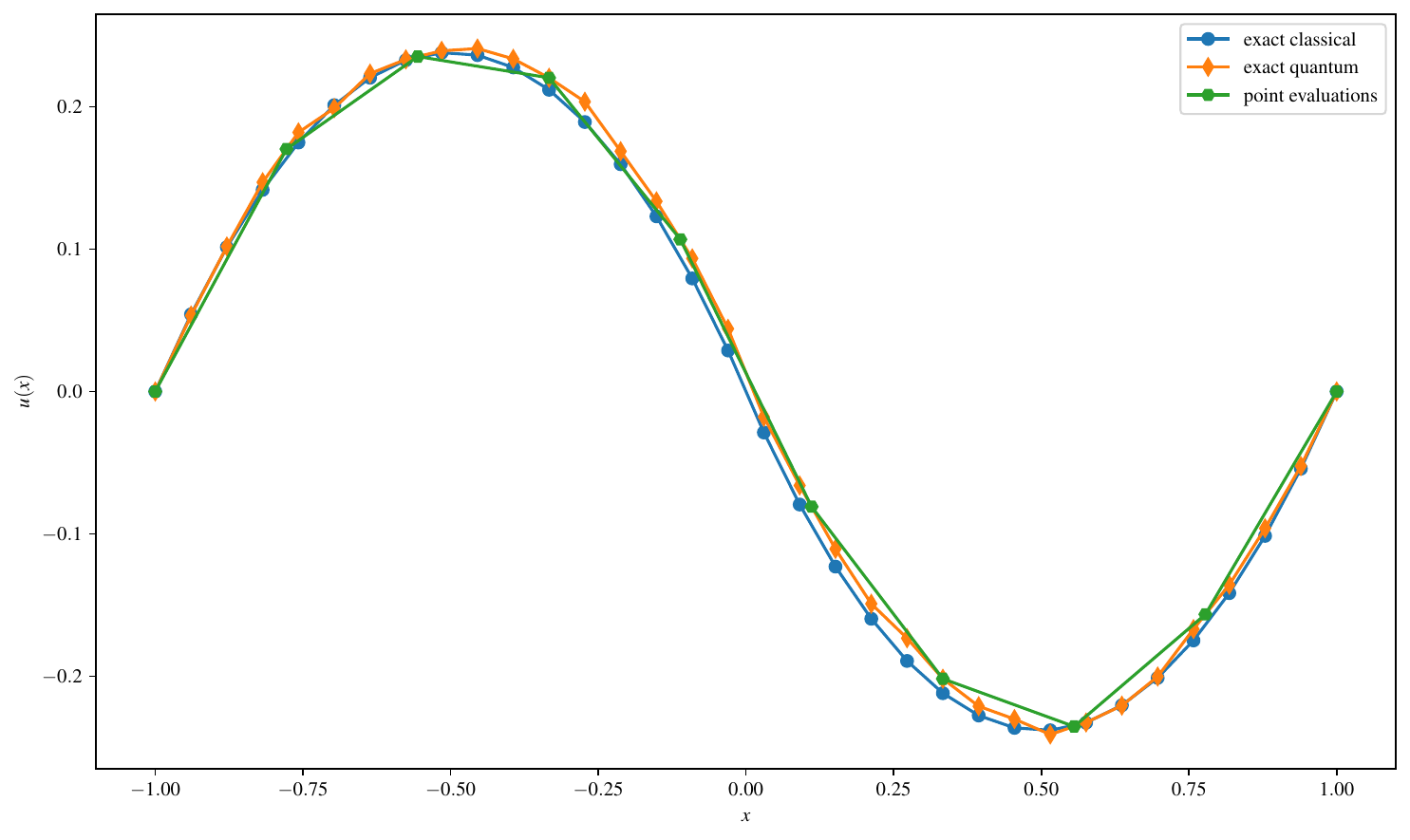}
  \caption{We plot the exact classical solution of the linear system of equations,
  the quantum state amplitudes on 5 qubits prepared with a variational solver as in \cite{Ali23} and
  the quantum solution evaluated at 10 points.}
  \label{fig:poisson}
\end{figure}

An illustrative example is provided by the Dirichlet problem for a second-order differential equation
\begin{alignat*}{2}
    -u''(x)&=f(x),&&\quad x\in(-1, 1),\\
    u(x)&=0,&&\quad x\in\{\pm 1\}.
\end{alignat*}
 The associated (rescaled) linear system, represented in matrix form,
\begin{align*}
\bs A\bs u=
\begin{pmatrix}
	2  & -1     & 0      & \dots & \\
	-1 & 2      & -1     & \dots & \\
	   &        & \ddots &       & \\
	   &        &        &     & -1\\ 
	   &        &        & -1    & 2 
\end{pmatrix}
\bs u=\bs f.
\end{align*}
can be solved using a variational quantum algorithm.
We use $\ket{\bs f}=H^{\otimes n-1}\otimes X\ket{0}$ for the RHS, which corresponds to a discontinuous
jump function. Details of all required circuits can be found in \cite{Ali23}.
The resulting quantum state $\ket{\bs u}$ approximates the normalized solution, and point evaluations of 
$u(x)$ can be efficiently computed and compared with the exact solution.
Figure \ref{fig:poisson} plots the exact normalized solution, the quantum approximation $\ket{\bs u}$ and the point evaluations $\frac{1}{\|\bs u\|}u(x)$.

\subsection{Regression}\label{sec:app:reg}
In the context of supervised learning, we are presented with a training set $\mc S_M:=\left\{(x_0, y_0),\ldots,(x_M,y_M)\right\}\in (X\times Y)^M$
consisting of independently and identically distributed samples from an unknown distribution $\mc P(X\times Y)$ and a hypothesis class $\mc H$.
The objective is to identify a model $h\in\mc H$, $h:X\times\Theta\rightarrow Y$, that minimizes the empirical risk
\begin{align*}
    R_M(h):=\frac{1}{M}\sum_i L(h(x_i, \theta), y_i),
\end{align*}
where $L:Y\times Y\rightarrow\R$ is a specified loss function.

In QML, hypothesis classes often derive from PQCs:
\begin{align*}
    \ket{\psi(x,\theta)}&:=U(x,\theta)\ket{0},\\
    h(x,\theta)&:=\tr(\ket{\psi(x,\theta)}\bra{\psi(x,\theta)}\mc M_\theta),\\
    \mc H&:=\{h(x,\theta):\theta\in\Theta\},
\end{align*}
with $\Theta$ being the parameter space.
Typically,  $U(x,\theta)$ incorporates data $x$ and parameters $\theta$ through rotational gates
as  in Figure \ref{fig:embeddings}, leading to Fourier series models in certain instances\footnote{This is the case if we
do not use feature-dependent multi-qubit gates for the data embedding, see also \cite{Schuld21}.}:
\begin{align}\label{eq:Fourier}
    h(x,\theta)=\sum_{\omega}c_\theta^\omega e^{-\mathrm{i}\omega\cdot x}.
\end{align}

\begin{figure}[ht!]
  \centering
  \includegraphics[width=0.9\textwidth, keepaspectratio]{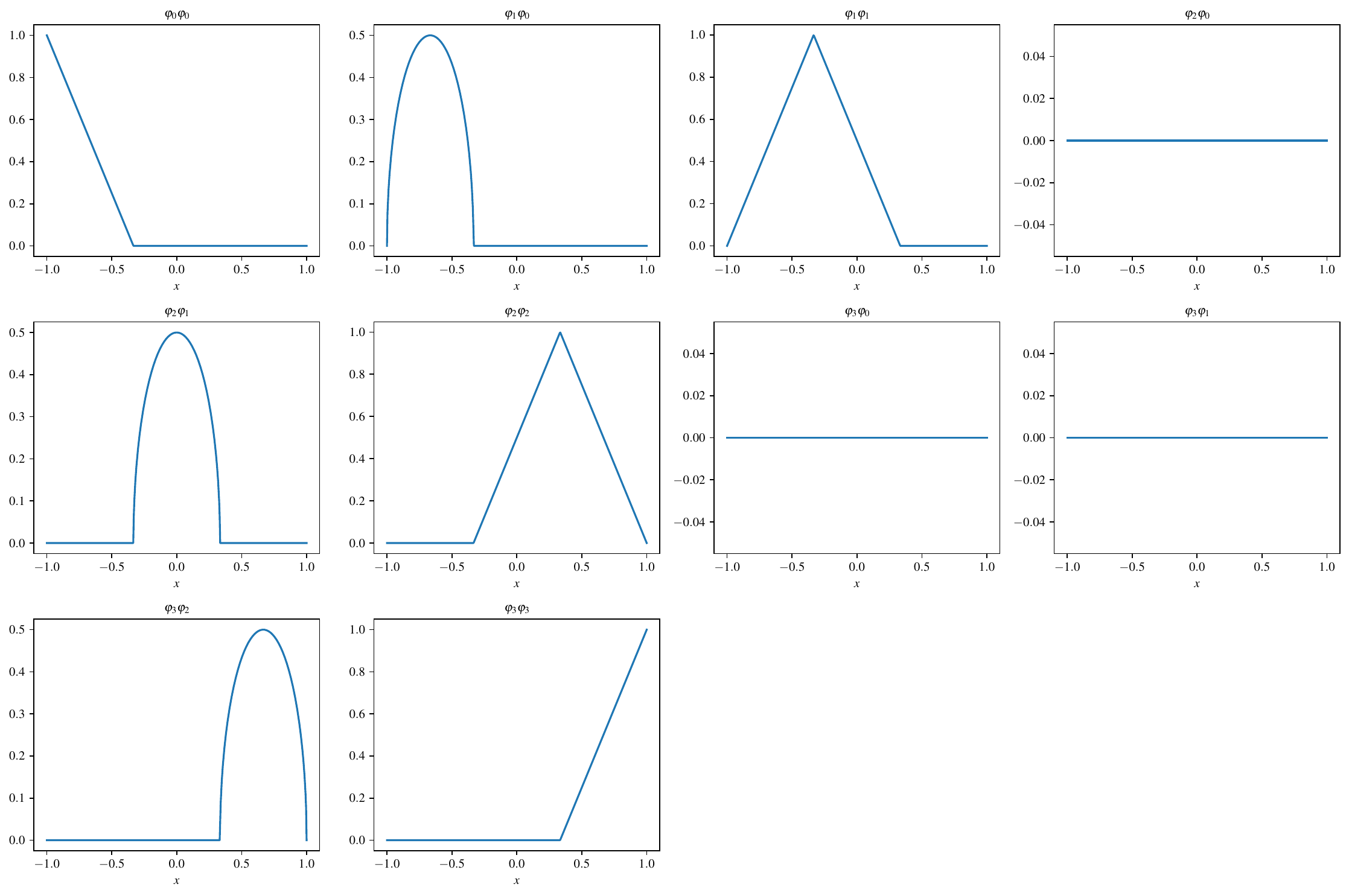}
  \caption{Plots of products of basis functions $\varphi_{\bar k}\varphi_{k}$, $\bar k\geq k$, on 2 qubits (10 products in total).}
  \label{fig:basis_funcs}
\end{figure}

Using PPTNQFE, we obtain the following expression for the model
\begin{align}\label{eq:model}
    h(x,\theta)&=\braket{\Phi(x)|U^\dag(\theta)\mc M_\theta U(\theta)|\Phi(x)}\notag\\
    &=(n_\Phi(x))^{-2}\sum_{\bar k, k}\varphi_{\bar k}(x)\varphi_k(x)
    \braket{\bar k|U^\dag(\theta)\mc M_\theta U(\theta)|k}\notag\\
    &=2(n_\Phi(x))^{-2}\sum_{\bar k\geq k}\varphi_{\bar k}(x)\varphi_k(x)
    \Re\braket{\bar k|U^\dag(\theta)\mc M_\theta U(\theta)|k},
\end{align}
for any unitary $U(\theta)$.
Note that for the $x$-dependence we obtain products $\varphi_{\bar k}(x)\varphi_k(x)$.
To include the original piecewise polynomials
from Section \ref{sec:pp} within this functional space, we set $\varphi_k:=\sqrt{\hat\varphi_k}$.
Due to property \eqref{req:unity} this automatically normalizes the basis vector so that $n_\Phi(x)=1$.
The entire set of functions $\varphi_{\bar k}\varphi_k$ is plotted in Figure \ref{fig:basis_funcs}.

It is crucial to recognize that these data embeddings are not inherently superior or inferior to conventional rotation-based embeddings;
they merely offer an alternative functional space that may better suit specific applications.
This diversity in embeddings introduces different inductive biases into the models.

\begin{figure}[htbp]
  \centering
  \begin{subfigure}[b]{0.49\textwidth}
    \includegraphics[width=\textwidth, keepaspectratio]{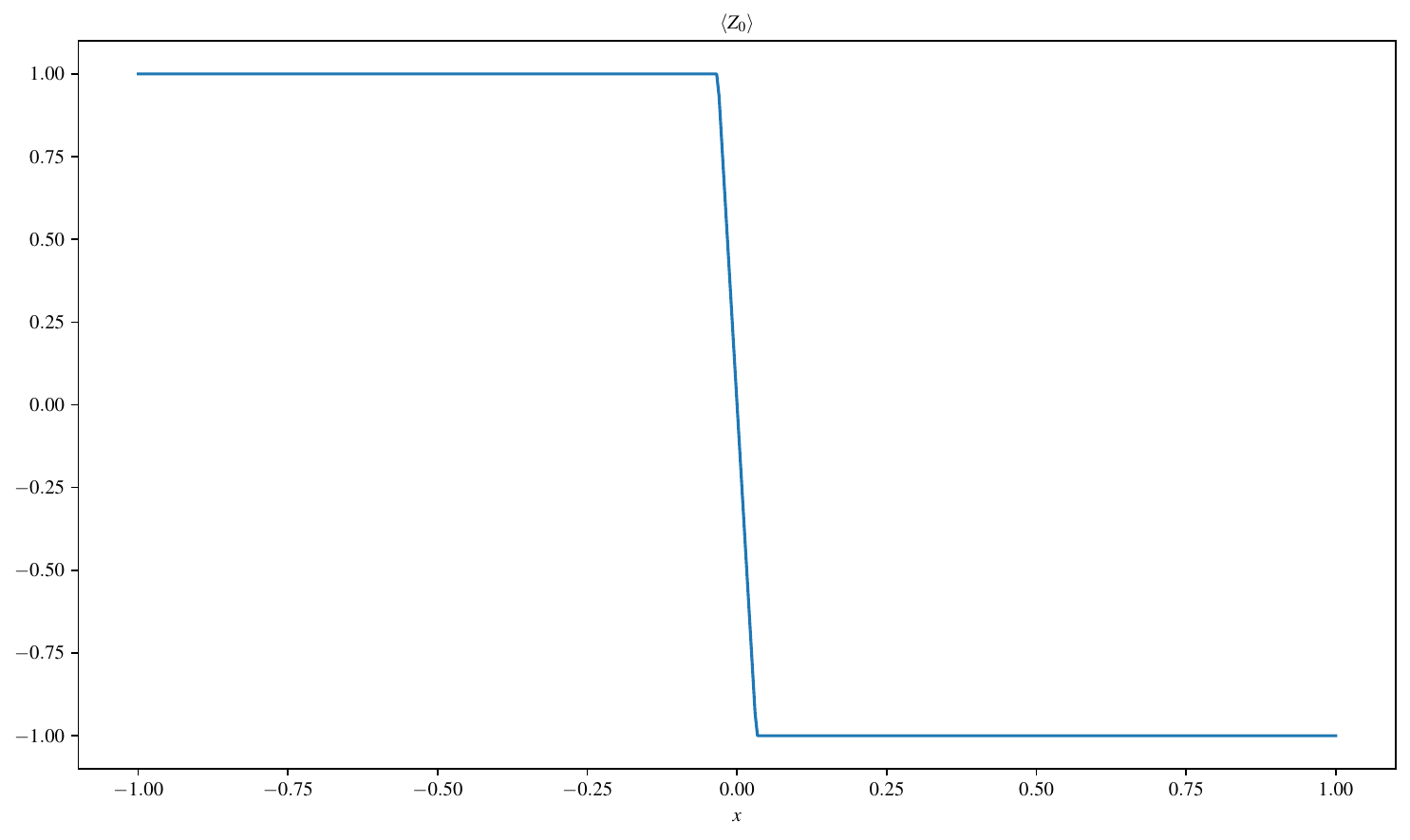}
  \end{subfigure}
  \begin{subfigure}[b]{0.49\textwidth}
    \includegraphics[width=\textwidth, keepaspectratio]{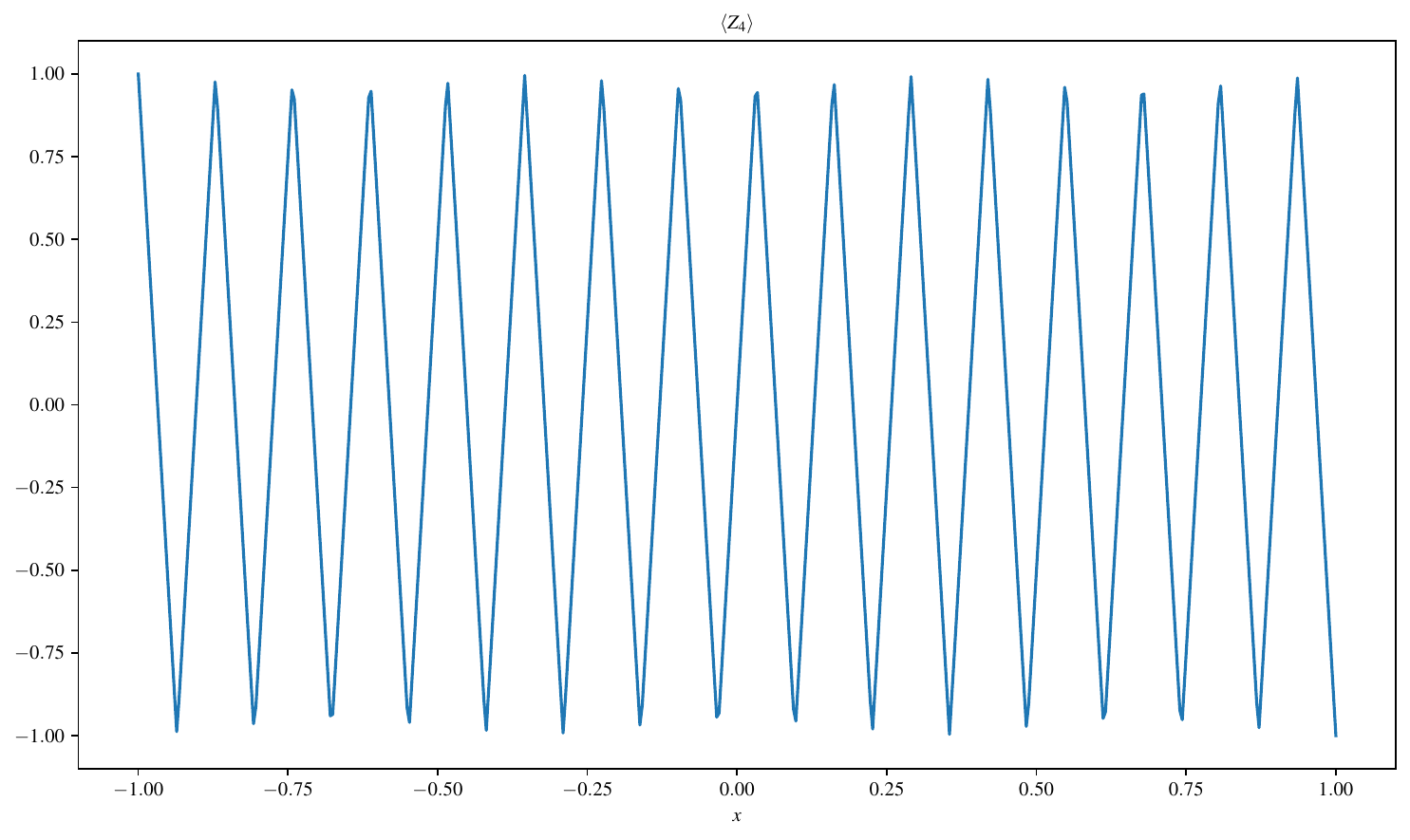}
  \end{subfigure}
  \begin{subfigure}[b]{0.49\textwidth}
    \includegraphics[width=\textwidth, keepaspectratio]{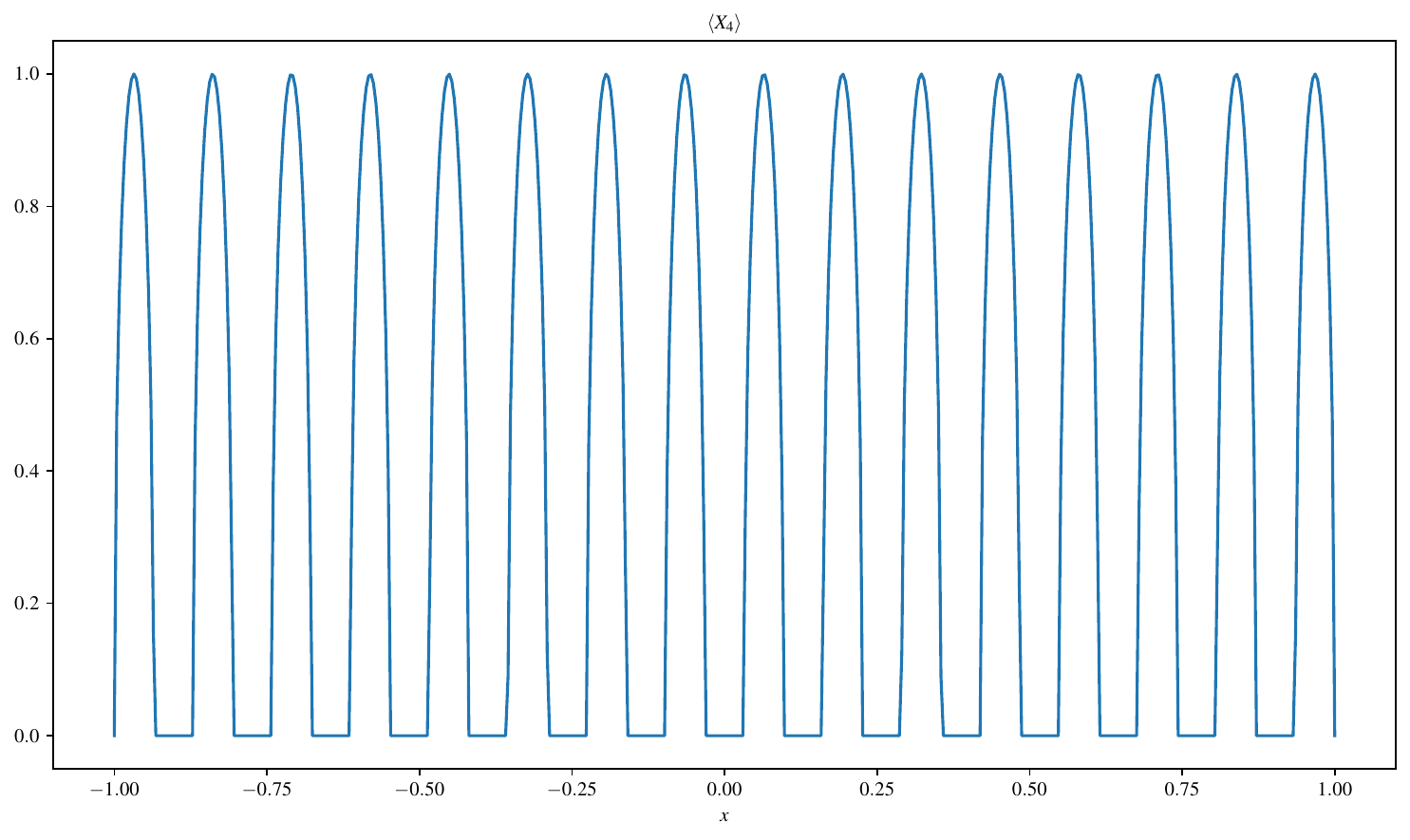}
  \end{subfigure}
  \caption{Plots of expectation values of different observables, applied
  immediately after the PPTNQFE embedding -- no variational layers or trainable parameters.}
  \label{fig:observables}
\end{figure}

To illustrate this point, consider Figure \ref{fig:observables}. These are plots of $h(x,\theta=0)$ for $x\in[-1, 1]$ and different observables $\mc M$, i.e.,
the data embedding contains no trainable layers and only parameter-independent observables.
From \eqref{eq:model}, we see that expectations values of
operators that are
diagonal in the computational $Z$-basis correspond
to combinations of piecewise polynomials
$\hat\varphi_k$ only. The qubit number on which we measure reflects the frequency of sign changes.
Vice versa, purely off-diagonal observables, like the Pauli-$X$, correspond to non-polynomial elements $\sqrt{\hat\varphi_{\bar k}\hat\varphi_k}$,
$\bar k\neq k$.
Similar statements hold for products of these observables.

The following training of quantum regression models was
performed on quantum state vector simulator with AMSGrad, using the MSE loss
on a training set of 50 equally spaced points.
More details can be found in the notebooks,
see Section \ref{sec:data}.

\begin{figure}[htbp]
  \centering
  \begin{subfigure}[b]{0.49\textwidth}
    \includegraphics[width=\textwidth, keepaspectratio]{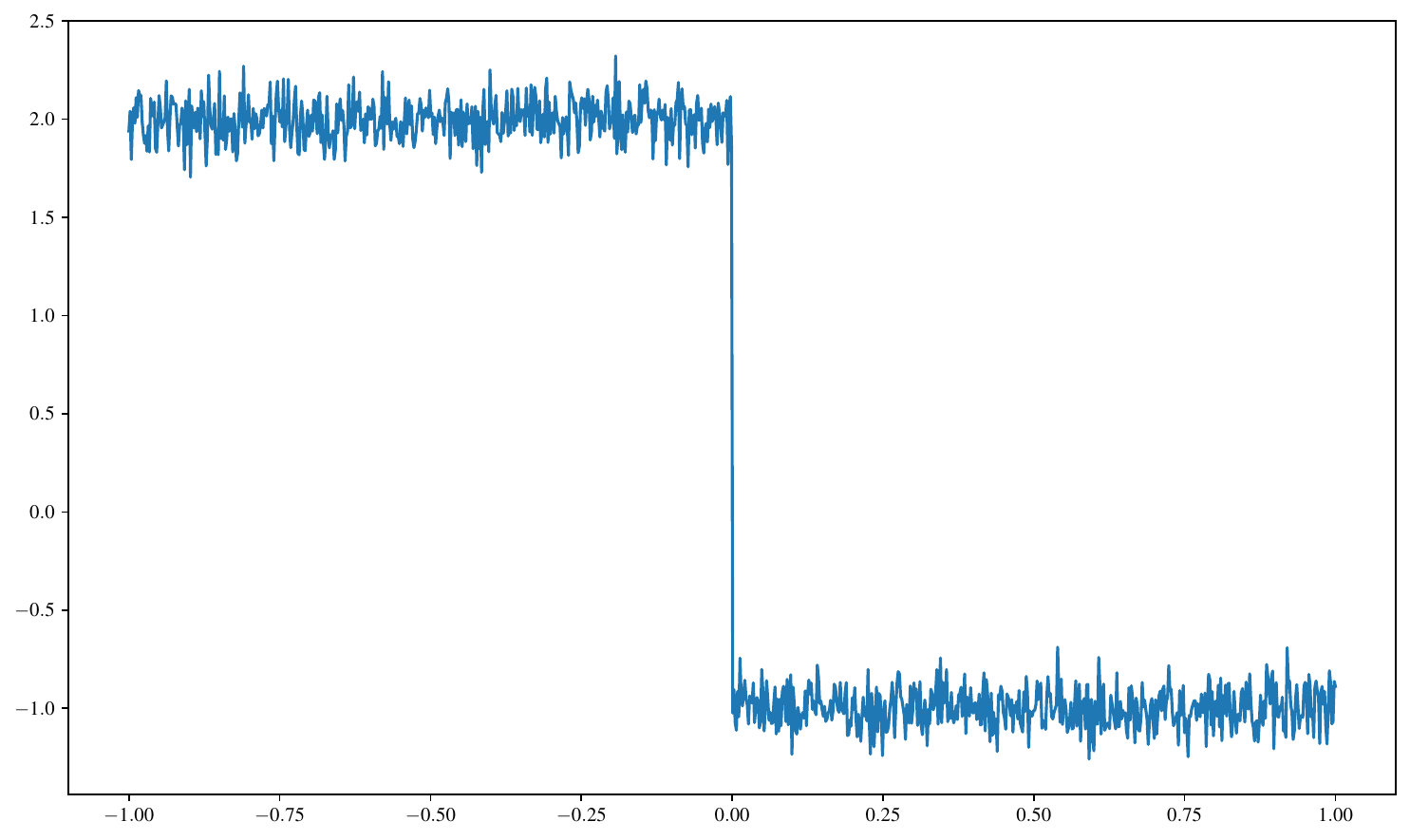}
  \end{subfigure}
  \begin{subfigure}[b]{0.49\textwidth}
    \includegraphics[width=\textwidth, keepaspectratio]{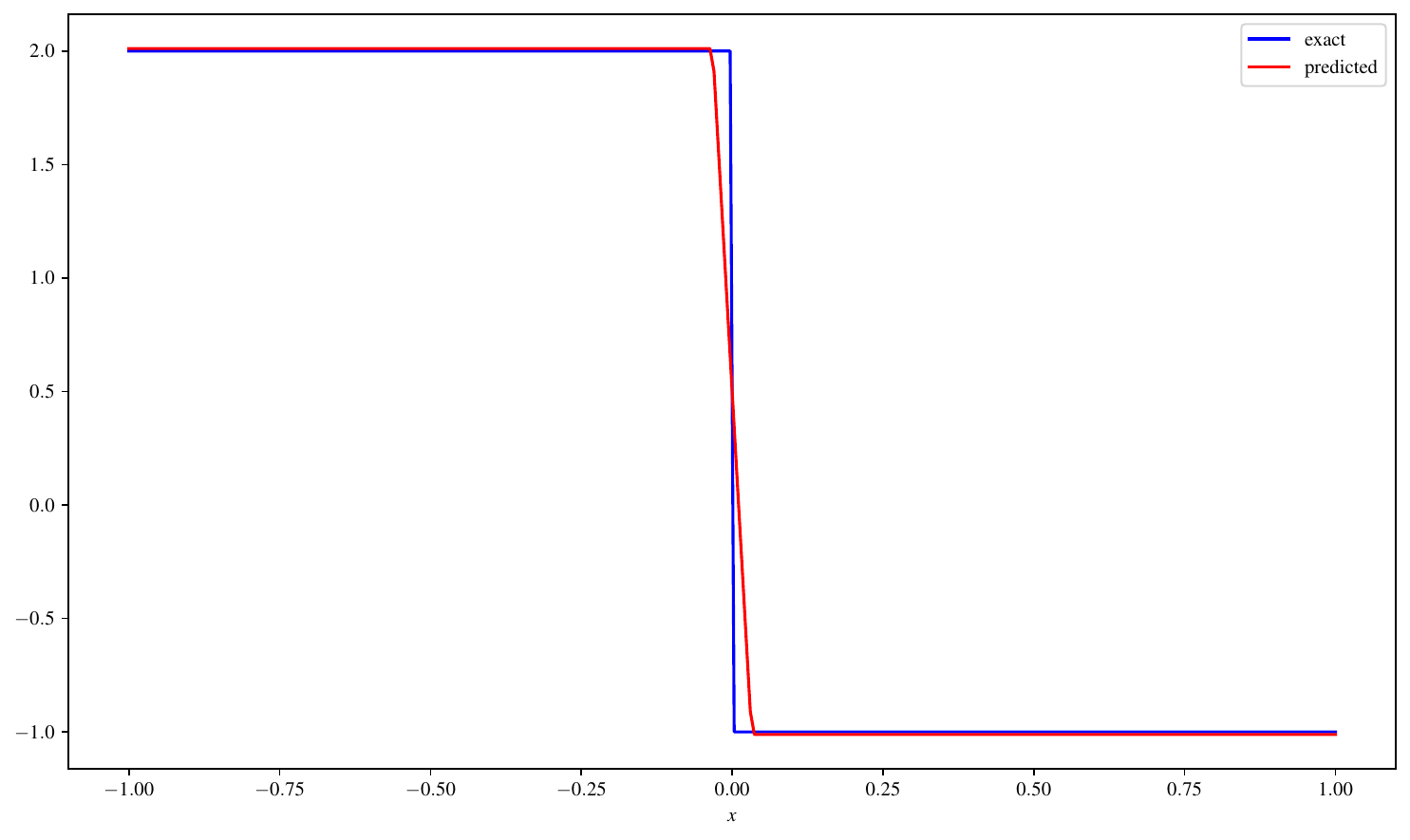}
  \end{subfigure}
  \begin{subfigure}[b]{0.49\textwidth}
    \includegraphics[width=\textwidth, keepaspectratio]{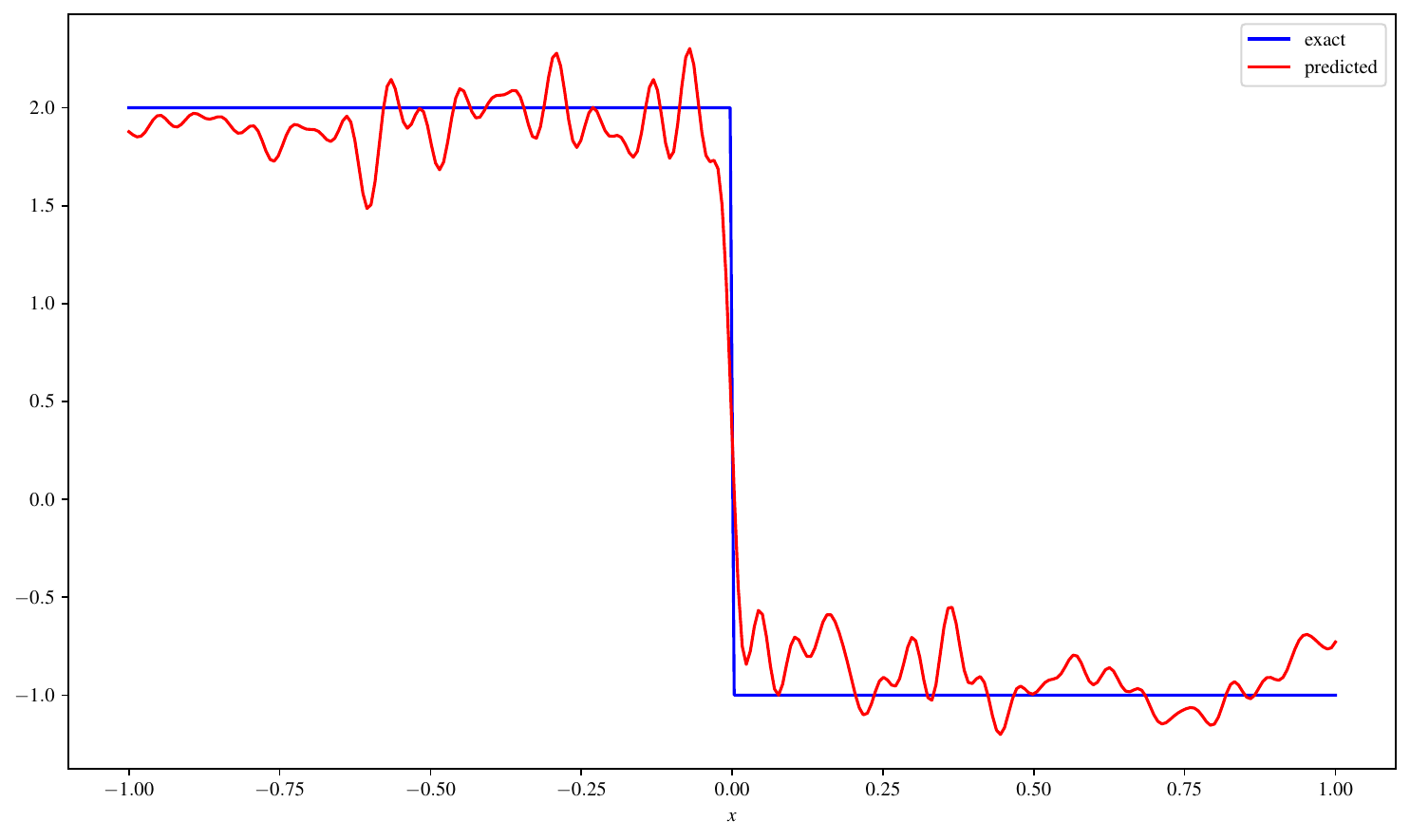}
  \end{subfigure}
  \caption{The training samples are based on the function top left. The PPTNQFE approximation is top right.
  The rotation-based model is on the bottom.}
  \label{fig:step}
\end{figure}

In Figure \ref{fig:step}, we approximate a discontinuous jump function with added Gaussian noise.
Traditional rotation-based embeddings will struggle with such functions, necessitating high frequencies to capture discontinuities.
On the contrary, PPTNQFE can achieve a satisfactory approximation with two trainable parameters.
For PPTNQFE, we use no variational layers and a parameter-dependent observable $\mc M_\theta=\theta_0I+\theta_1Z_0$,
where $Z_0$ is the $Z$-operator applied to the $0$'th qubit.
For the rotation-based model, we use exponential feature encoding as in Figure \ref{fig:embeddings},
the same observable and 3 variational layers, resulting in a total of 89 trainable parameters.

Next, consider a $\sin$ wave perturbed by Gaussian noise.
For such a function, a rotation-based model with a limited frequency spectrum would generally provide an adequate approximation.
However, to demonstrate overfitting,
we employ exponential encoding in the rotation-based embedding, as illustrated in Figure \ref{fig:embeddings}. 
We attempt fitting the rotation-based model on 5 qubits with a
\begin{itemize}
    \item a single variational layer consisting of universal single-qubit gates and CNOTs,
    combined with 5 trainable observables $\theta_k Z_k$ ($Z_k$ is the $Z$-operator applied to qubit $k$), and a constant bias term, culminating in a total of 45 trainable parameters;
    \item and an expanded setup with 3 variational layers, increasing the total count of trainable parameters to 93.
\end{itemize}

\begin{figure}[htbp]
  \centering
  \begin{subfigure}[b]{0.49\textwidth}
    \includegraphics[width=\textwidth, keepaspectratio]{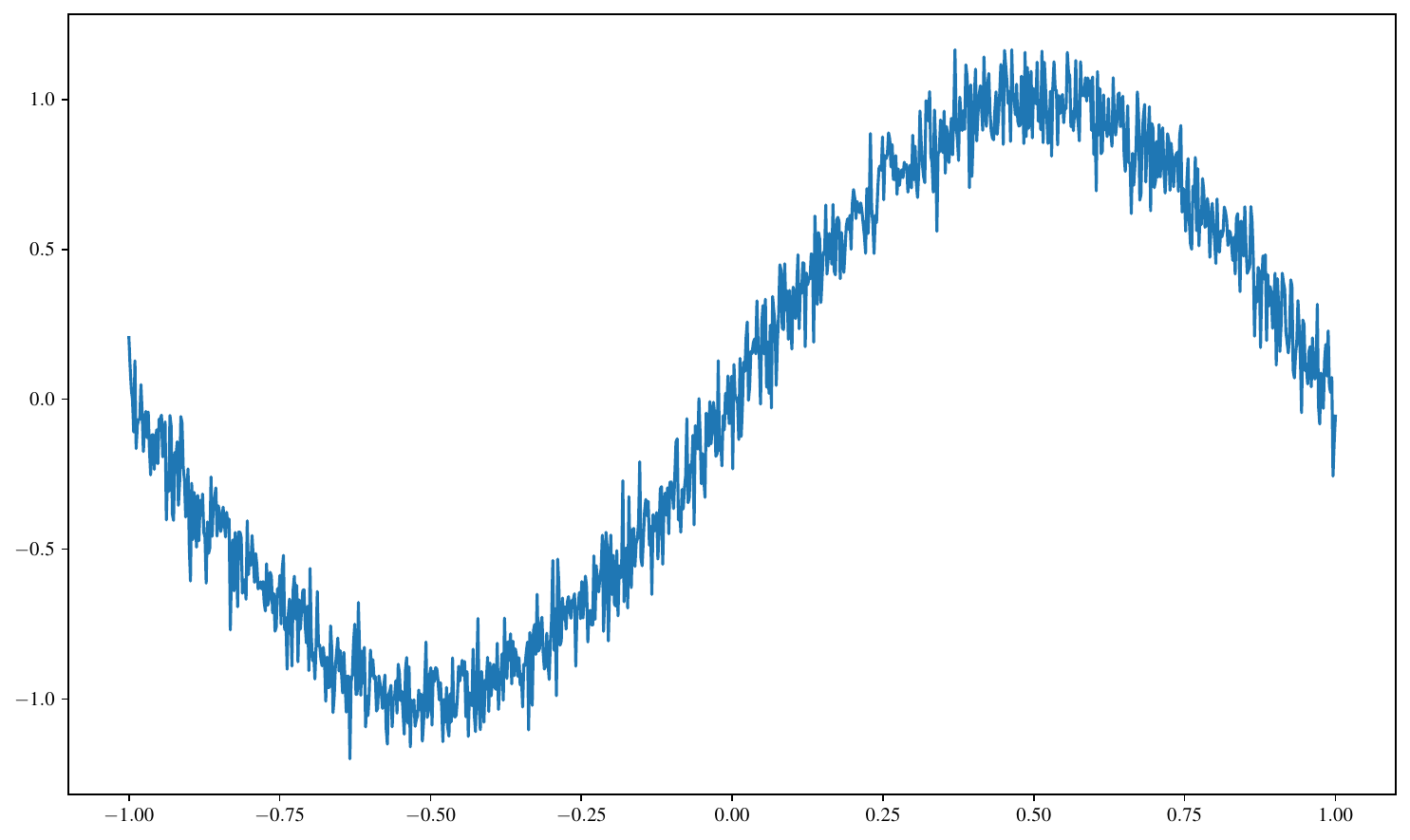}
  \end{subfigure}
  \begin{subfigure}[b]{0.49\textwidth}
    \includegraphics[width=\textwidth, keepaspectratio]{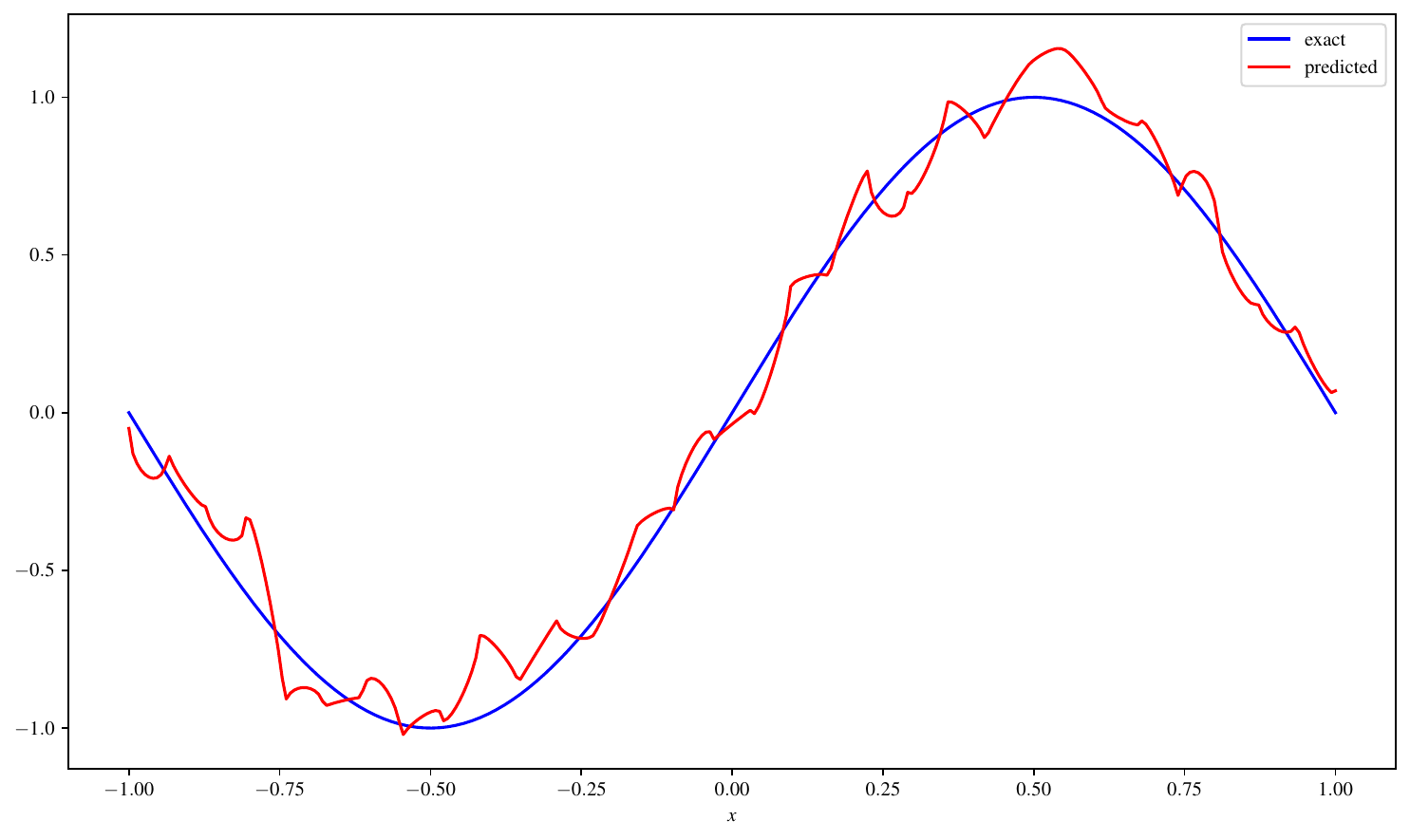}
  \end{subfigure}
  \begin{subfigure}[b]{0.49\textwidth}
    \includegraphics[width=\textwidth, keepaspectratio]{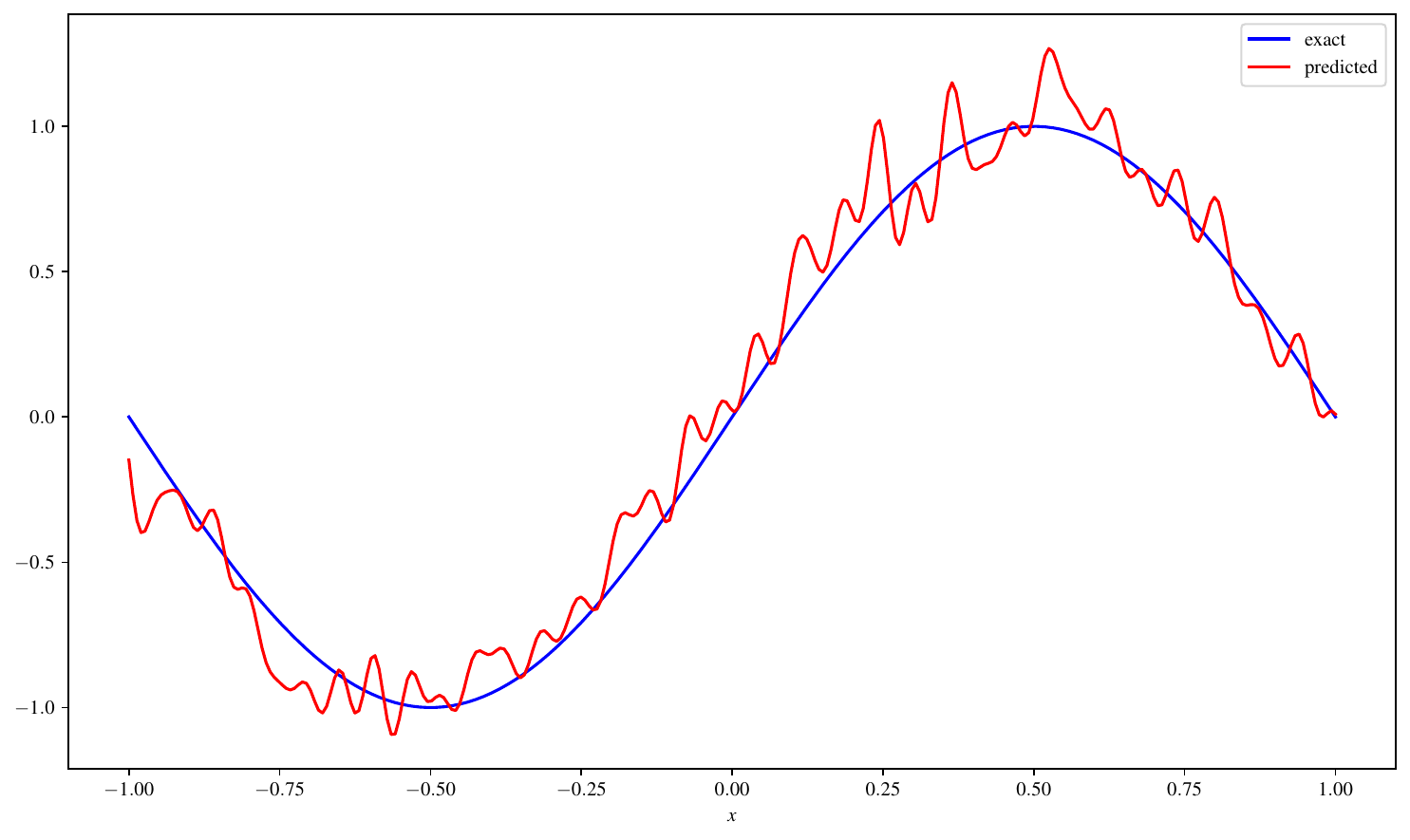}
  \end{subfigure}
  \begin{subfigure}[b]{0.49\textwidth}
    \includegraphics[width=\textwidth, keepaspectratio]{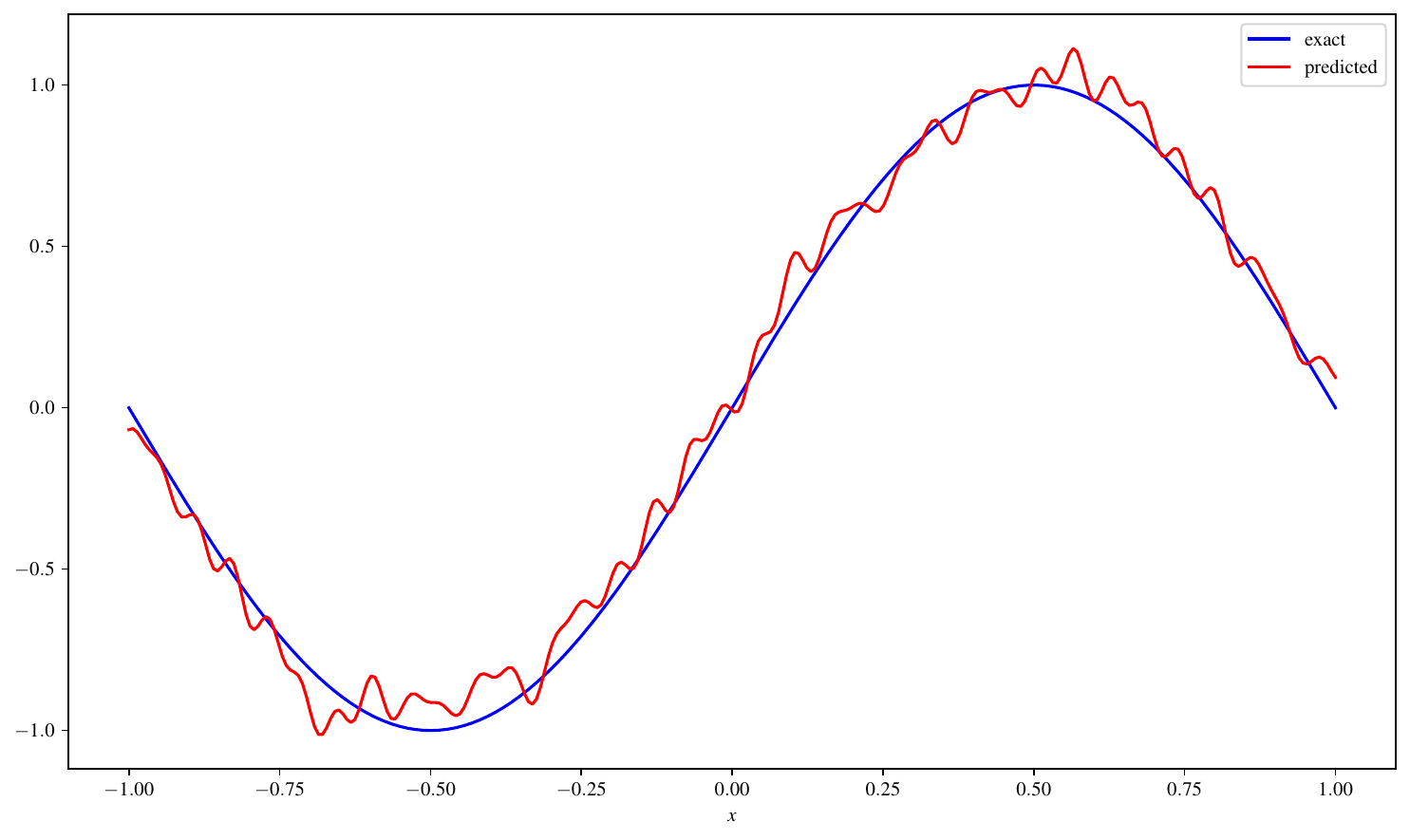}
  \end{subfigure}  
  \caption{The training samples are based on the function top left.
  The PPTNQFE approximation with an MPS variational layer (93 parameters) is top right.
  The rotation-based model with 3 variational layers (93 parameters) is bottom left.
  The rotation-based model with 1 variational layers (45 parameters) is bottom right.}
  \label{fig:sine}
\end{figure}

For PPTNQFE, we use an MPS-structured variational layer, similar to Figure \ref{fig:embeddings},
but with three 2-qubit gates instead of a single layer.
The rationale for employing an MPS-structured variational layer is based on the idea that the $\sin$ function can be represented as a rank-2 MPS, as detailed in
\cite{Oseledets13}. By using a variational layer with a similar structure
to the PPTNQFE embedding layer, the aim is to discover a unitary transformation that effectively maps back to the computational basis.
This method has shown improved outcomes compared to the use of generic variational layers, although it remains a heuristic approach at this stage.
We use observables similar to those in the rotation-based model, $\mc M_\theta = \theta_0 + \sum_k \theta_{k+1}Z_k$, resulting in 93 trainable parameters in total.
The outcomes are presented in Figure \ref{fig:sine}.

Both rotation-based models overfit noise, with the simpler one performing slightly better.
This is mainly due to the accessible spectrum being too large and no bias towards the correct solution.
The number of parameters was deliberately taken to be larger than necessary to demonstrate noise overfitting.
Similarly, PPTNQFE with 93 parameters does not perform as well as for the jump function,
with square root cusps introduced due to the added variational
layers in the model. The occurrence of non-polynomial terms
was explained in Figure \ref{fig:observables}.

Given a budget of $n$ qubits and using PPTNQFE, we can encode $2^n$ piecewise polynomials and, taking into account off-diagonal terms,
$2^{2n-1}+2^{n-1}$ non-polynomial terms.
Measuring simple observables such as $Z_0$ means,
the final approximation is a linear combination
of all piecewise polynomials.
If we want a local approximation, addressing each piecewise polynomial $\hat\varphi_k$ individually,
doing so naively would require measuring $$\mc M_\theta:=\sum_k\theta_k\ket{k}\bra{k},$$
with $2^n$ trainable parameters.
This is similar to the rotation-based models from \eqref{eq:Fourier},
where a limited number of parameters means we only have access to a small subset of
the Fourier coefficients $\{c_\theta^\omega\}_\omega$.
There is no ``free lunch'' in either case, and, despite the ambient Hilbert space being exponentially large, a good hypothesis class $\mc H$ for the
given learning problem will require choosing the right structure for the trainable unitaries and observables.

\begin{figure}[htbp]
  \centering
  \begin{subfigure}[b]{0.49\textwidth}
    \includegraphics[width=\textwidth, keepaspectratio]{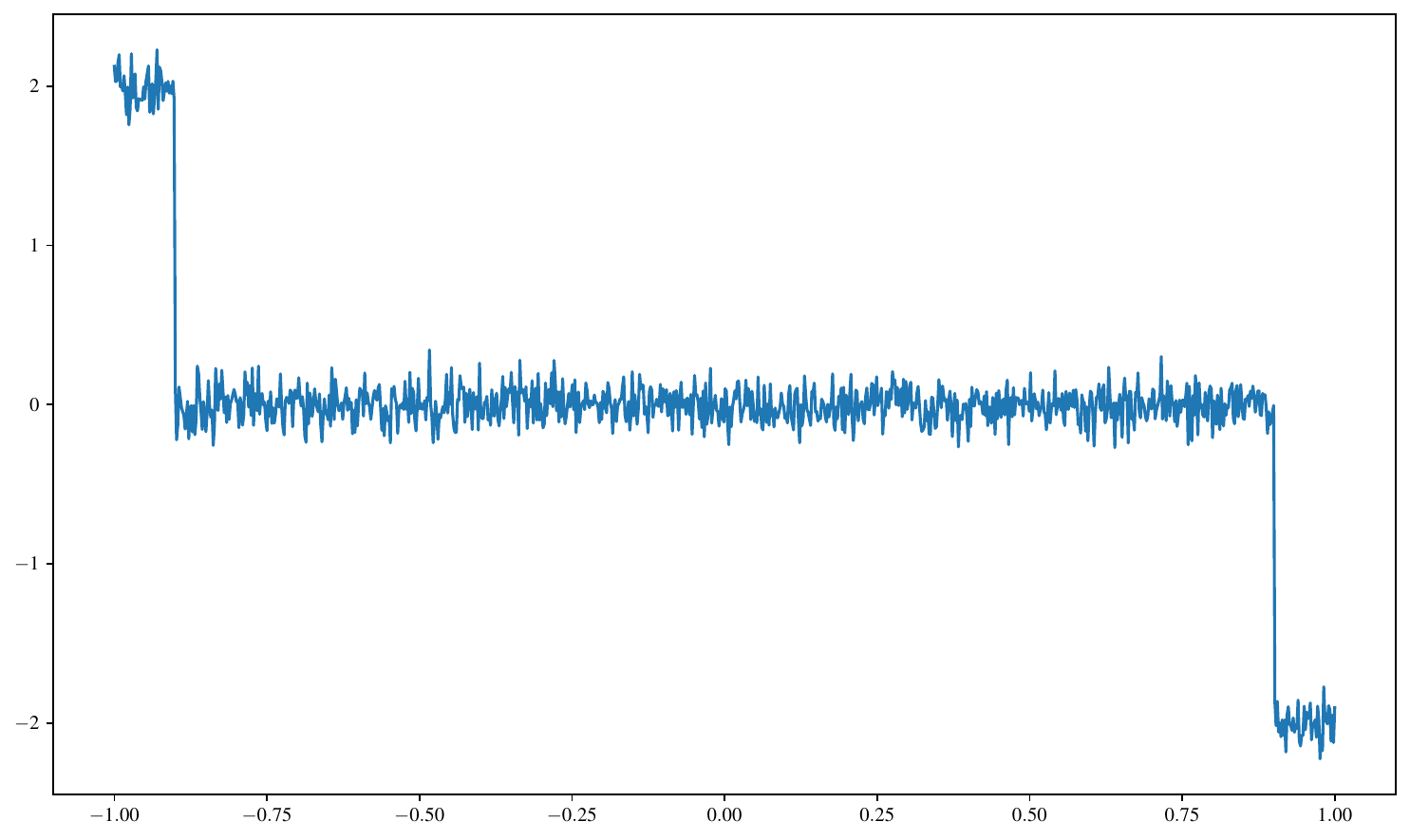}
  \end{subfigure}
  \begin{subfigure}[b]{0.49\textwidth}
    \includegraphics[width=\textwidth, keepaspectratio]{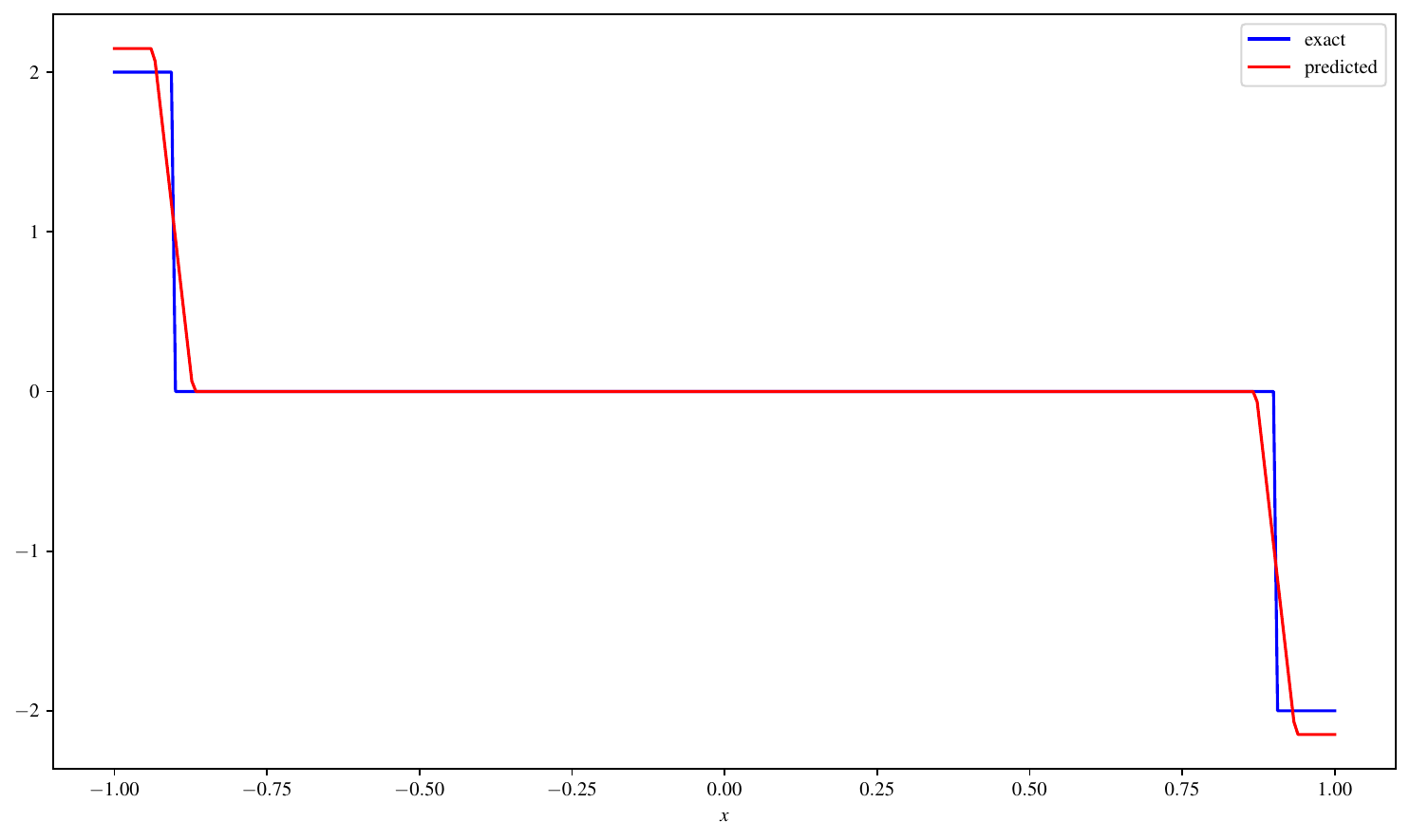}
  \end{subfigure}
  \caption{The training samples are based on the function left.
  The PPTNQFE model appends a non-trainable layer based on an analytically computed unitary and
  measures $\mc M_\theta = \theta Z_0$.}
  \label{fig:multistep}
\end{figure}

As an example, consider the observable $\mc M_\theta:=\theta Z_0$,
and the target function from Figure \ref{fig:multistep} with 2 discontinuous jumps.
Clearly, a sufficiently good approximation would only take the first two and the last two piecewise polynomials
$\hat\varphi_k$, $k=0, 1, 2^n-2, 2^n-1$.
From \eqref{eq:model}, we see that this would require a unitary that rotates into a basis
in which the observable $Z_0$ has diagonal entries $1, 1, 0,\ldots, 0, -1, -1$ and arbitrary, but purely complex-valued,
off-diagonal entries.
As can be seen in Figure \ref{fig:multistep}, this is indeed achievable. Here, the unitary was obtained analytically by direct optimization,
see Section \ref{sec:data}. More generally, problem-specific, efficient quantum circuit designs is an interesting open question.

\subsection{Classification}\label{sec:classify}
For the final example, we consider classification with quantum kernels. As pointed out in \cite{schuld2021}, supervised QML methods without data re-uploading are kernel methods. Thus, we can determine the optimal solution by minimizing the hinge loss over a training set through inverting the kernel matrix. If \( x \mapsto \rho(x) \) is the embedding map, the kernel map is given by

\[
    \kappa(x, x') = \tr[\rho(x')\rho(x)],
\]

and the prediction for any new point \( x^* \) is computed as

\[
    f(x^*) = \sum_{m=1}^M \alpha_m \kappa(x^*, x_m),
\]

where \( M \) is the number of training samples and the coefficients
\( \alpha_m \) are determined during training.

\begin{figure}[htbp]
  \centering
  \begin{subfigure}[b]{0.49\textwidth}
    \includegraphics[width=\textwidth, keepaspectratio]{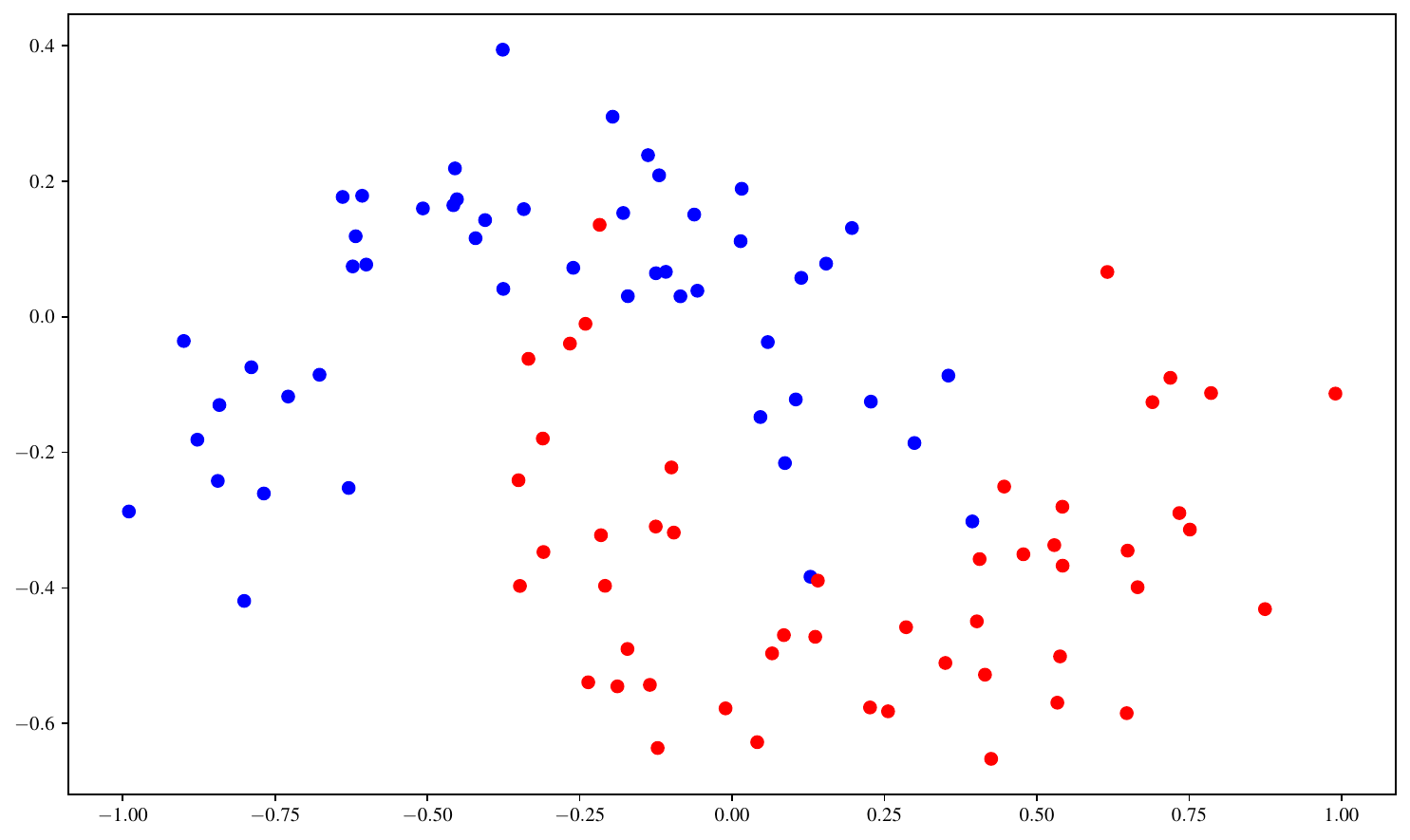}
  \end{subfigure}
  \begin{subfigure}[b]{0.49\textwidth}
    \includegraphics[width=\textwidth, keepaspectratio]{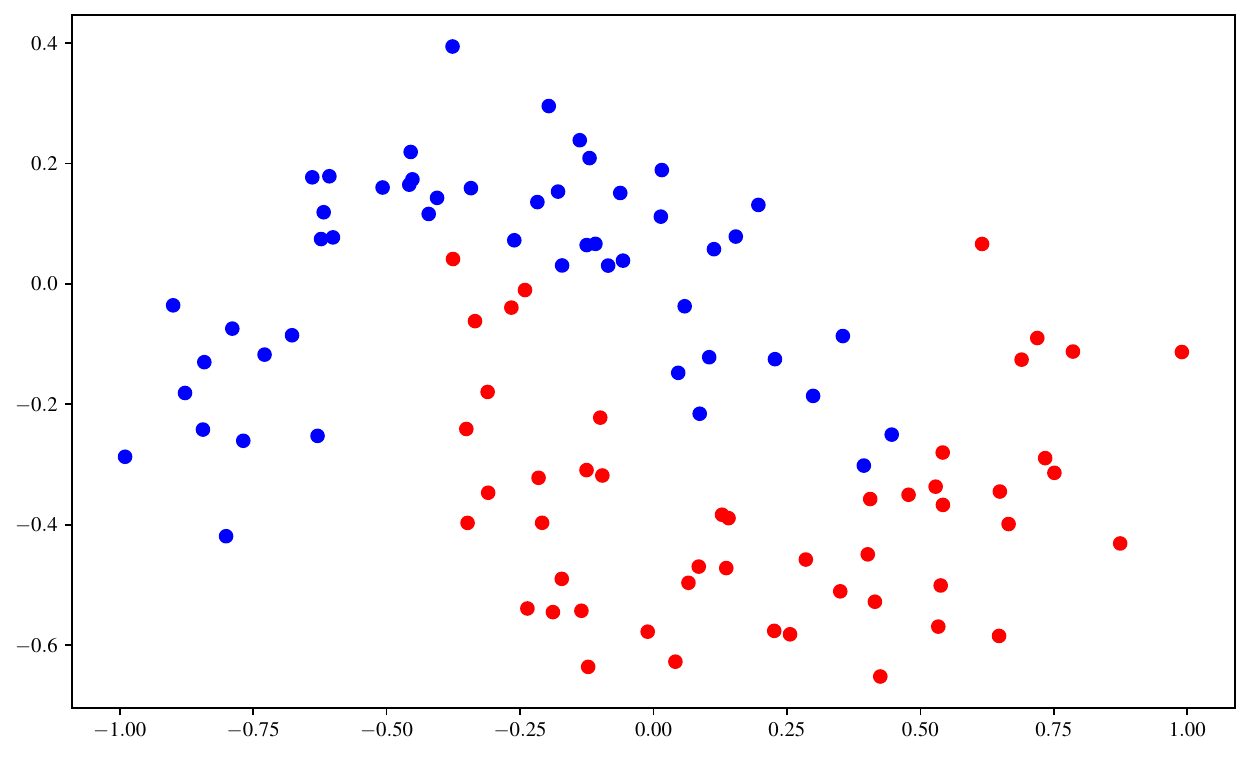}
  \end{subfigure}
  \begin{subfigure}[b]{0.49\textwidth}
    \includegraphics[width=\textwidth, keepaspectratio]{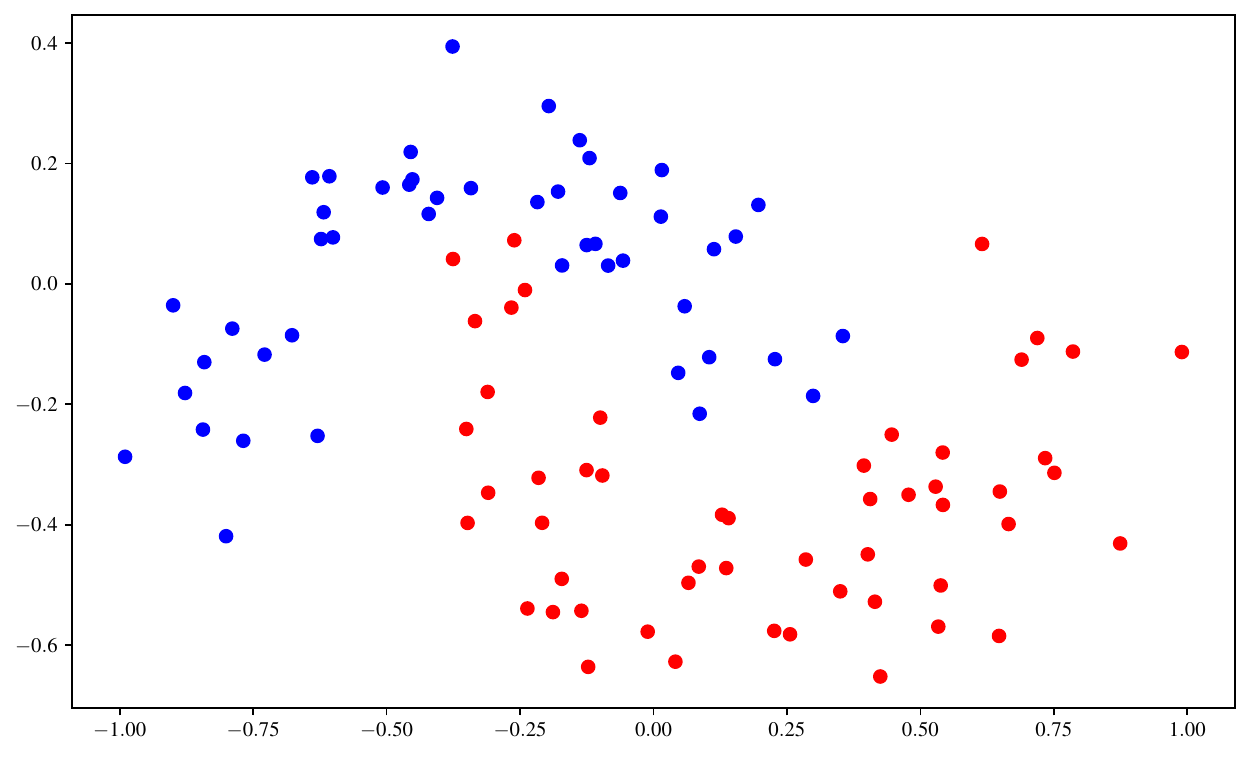}
  \end{subfigure}
  \caption{Top left: full dataset of two half moons
  representing two classes with noise.
  The training set for both quantum kernels was chosen as 80\% of the full dataset,
  randomly selected.
  Top right: classification results on the full dataset for the angle-based
  embedding after training.
  Bottom: classification results on the full dataset for PPTNQFE
  after training.}
  \label{fig:moons}
\end{figure}

\begin{figure}[htbp]
  \centering
  \begin{subfigure}[b]{0.49\textwidth}
    \includegraphics[width=\textwidth, keepaspectratio]{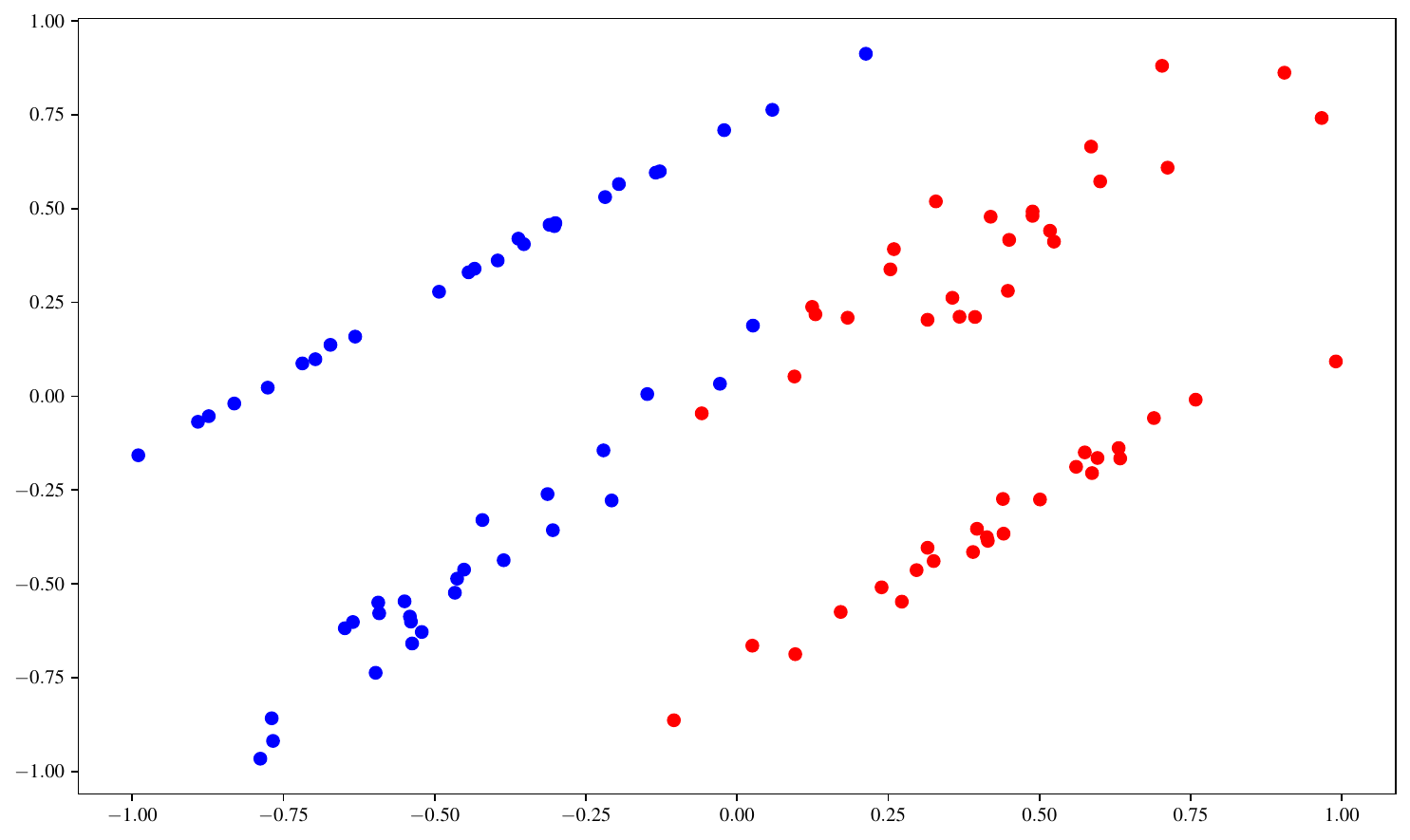}
  \end{subfigure}
  \begin{subfigure}[b]{0.49\textwidth}
    \includegraphics[width=\textwidth, keepaspectratio]{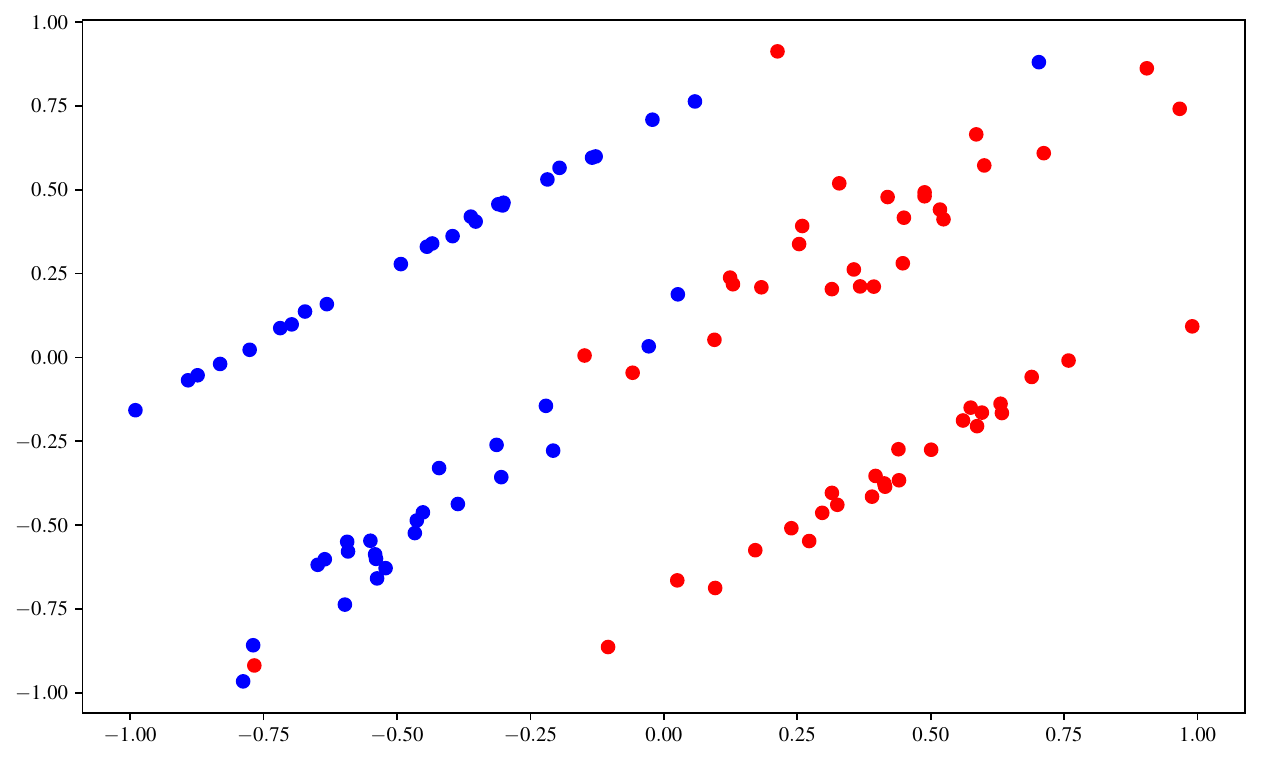}
  \end{subfigure}
  \begin{subfigure}[b]{0.49\textwidth}
    \includegraphics[width=\textwidth, keepaspectratio]{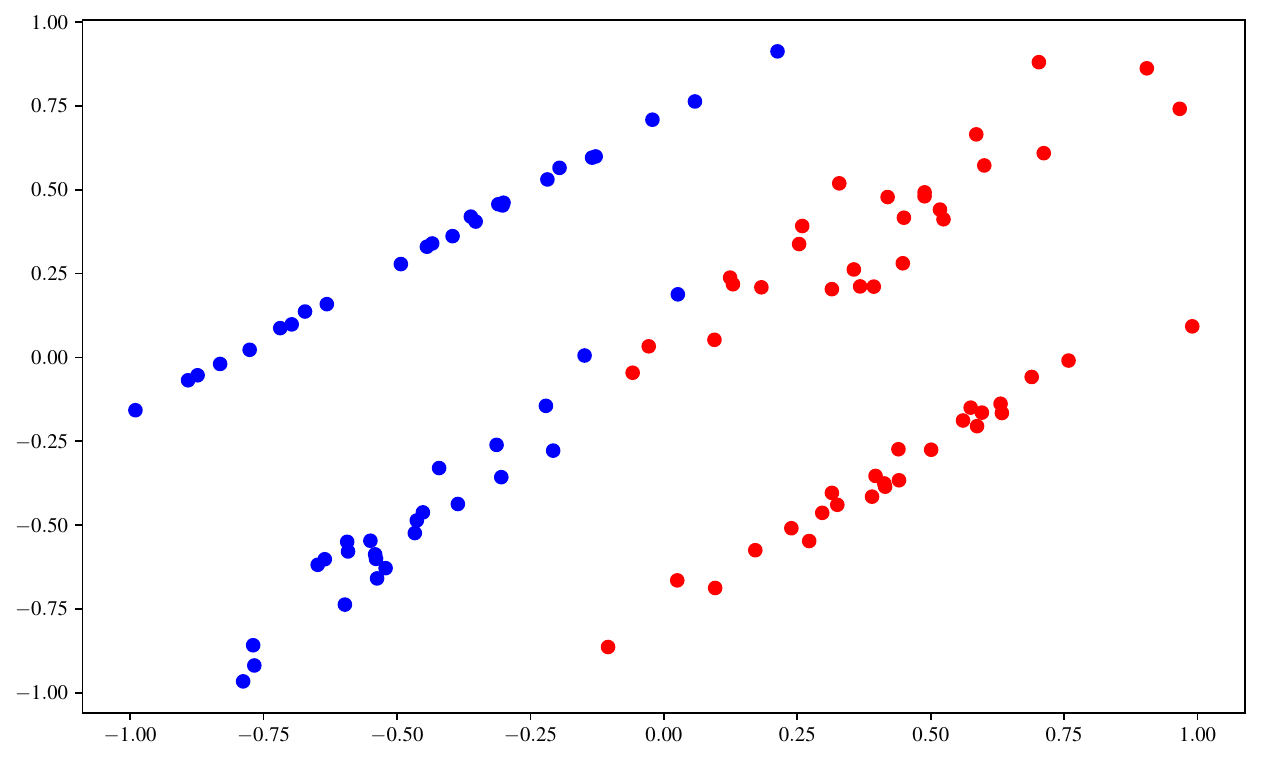}
  \end{subfigure}
  \caption{Top left: full dataset of three lines
  representing two classes with noise.
  The training set for both quantum kernels was chosen as 80\% of the full dataset,
  randomly selected.
  Top right: classification results on the full dataset for the angle-based
  embedding after training.
  Bottom: classification results on the full dataset for PPTNQFE
  after training.}
  \label{fig:lines}
\end{figure}

We apply this method to both the angle-based embedding and PPTNQFE, and test the quantum kernel classification on two synthetic datasets, as shown in Figures \ref{fig:moons} and \ref{fig:lines}. For a fair comparison, we use 6 qubits for both models. For the angle-based embedding, this corresponds to a Fourier spectrum of 13 different frequencies per feature (169 in total), and for PPTNQFE, this corresponds to 8 hat functions for each feature (64 in total).

Both models perform similarly well on the two half moons dataset. Some points are misclassified due to noise, which places some points of different classes very close to each other.

For the line-like clusters in Figure \ref{fig:lines}, a simple linear model should already perform well. However, the angle-based embedding misclassifies some points that are far from the clusters due to the periodic nature of the embedding and the presence of high-frequency components in the model. On the other hand, PPTNQFE does not suffer from this issue and only misclassifies one point that is close to the boundary between the two clusters.

\section{Conclusions}\label{sec:concl}
In this work, we introduced a methodology to embed continuous variables in quantum circuits using piecewise polynomial features.
This approach involves representing a piecewise polynomial basis as a low-rank TN, which is then transformed into a quantum circuit.
The conversion process results in a quantum circuit where the number of qubits for the converted TN quantum gates
scales logarithmically with the rank of the TN, making the embeddings potentially efficient for low-rank TNs with the number of gates growing linearly with the qubit count.

However, converting TN gates to those native to quantum hardware may introduce significant overhead.
Additionally, the effects of statistical and hardware-induced noise on these embeddings were not explored in this work.

We explored three applications of this method.
First, PPTNQFE allows efficient point evaluations of discretized solutions to PDEs, enhancing the appeal of quantum algorithms for solving PDEs by addressing data read-out challenges.
Second, we demonstrated the utility of PPTNQFE in function regression, particularly highlighting its aptitude for modeling functions with localized features, such as jump discontinuities.
Third, we showed that PPTNQFE can easily identify non-linear decision boundaries, similar to angle-based embeddings, while avoiding some of the typical issues associated with the periodic Fourier nature of angle-based embeddings.
An interesting open question is how to design problem-specific trainable circuits suited for approximating more complex
dependencies with local effects.

The broader objective of our research is to widen the utility of quantum models by introducing a novel approach to feature embeddings that takes advantage of TN representations.
This not only facilitates the embedding of piecewise polynomials, but also opens avenues for incorporating various other function types into quantum models.
The rich existing literature on the theoretical and numerical aspects of TN representations could provide valuable insights, contributing to a more profound understanding of the potential of QML.
We aim to unveil the inner mechanics of quantum models in a transparent manner, focusing on comprehending how design choices impact their functionality, the scope of problems they are suited for, and exploring their potential benefits, if any.

\section{Data Availability}\label{sec:data}
The source code for all the numerical experiments presented here can be found at \url{https://github.com/MazenAli/PPTNQFE_Examples}.
Running the notebooks requires installing the package \url{https://github.com/MazenAli/QuLearn}.

\section{Acknowledgments}
We thank BMWK for the financial support provided under the \emph{EniQmA} project and the Competence Center for Quantum Computing Rhineland-Palatinate for the financial support provided under the \emph{AnQuC-3} project.

\bibliographystyle{unsrtnat}
\bibliography{references}  

\begin{thebibliography}{54}
\providecommand{\natexlab}[1]{#1}
\providecommand{\url}[1]{\texttt{#1}}
\expandafter\ifx\csname urlstyle\endcsname\relax
  \providecommand{\doi}[1]{doi: #1}\else
  \providecommand{\doi}{doi: \begingroup \urlstyle{rm}\Url}\fi

\bibitem[Huang et~al.(2021{\natexlab{a}})Huang, Kueng, and Preskill]{Preskill21}
Hsin-Yuan Huang, Richard Kueng, and John Preskill.
\newblock Information-theoretic bounds on quantum advantage in machine learning.
\newblock \emph{Phys. Rev. Lett.}, 126:\penalty0 190505, May 2021{\natexlab{a}}.
\newblock \doi{10.1103/PhysRevLett.126.190505}.
\newblock URL \url{https://link.aps.org/doi/10.1103/PhysRevLett.126.190505}.

\bibitem[Huang et~al.(2022)Huang, Broughton, Cotler, Chen, Li, Mohseni, Neven, Babbush, Kueng, Preskill, and McClean]{Preskill22}
Hsin-Yuan Huang, Michael Broughton, Jordan Cotler, Sitan Chen, Jerry Li, Masoud Mohseni, Hartmut Neven, Ryan Babbush, Richard Kueng, John Preskill, and Jarrod~R. McClean.
\newblock Quantum advantage in learning from experiments.
\newblock \emph{Science}, 376\penalty0 (6598):\penalty0 1182--1186, 2022.
\newblock \doi{10.1126/science.abn7293}.
\newblock URL \url{https://www.science.org/doi/abs/10.1126/science.abn7293}.

\bibitem[Lee et~al.(2023)Lee, Lee, Zhai, Tong, Dalzell, Kumar, Helms, Gray, Cui, Liu, Kastoryano, Babbush, Preskill, Reichman, Campbell, Valeev, Lin, and Chan]{Preskill23}
Seunghoon Lee, Joonho Lee, Huanchen Zhai, Yu~Tong, Alexander~M. Dalzell, Ashutosh Kumar, Phillip Helms, Johnnie Gray, Zhi-Hao Cui, Wenyuan Liu, Michael Kastoryano, Ryan Babbush, John Preskill, David~R. Reichman, Earl~T. Campbell, Edward~F. Valeev, Lin Lin, and Garnet Kin-Lic Chan.
\newblock Evaluating the evidence for exponential quantum advantage in ground-state quantum chemistry.
\newblock \emph{Nature Communications}, 14\penalty0 (1):\penalty0 1952, April 2023.
\newblock ISSN 2041-1723.
\newblock \doi{10.1038/s41467-023-37587-6}.
\newblock URL \url{https://doi.org/10.1038/s41467-023-37587-6}.

\bibitem[Abbas et~al.(2023)Abbas, King, Huang, Huggins, Movassagh, Gilboa, and McClean]{Abbas2023}
Amira Abbas, Robbie King, Hsin-Yuan Huang, William~J. Huggins, Ramis Movassagh, Dar Gilboa, and Jarrod~R. McClean.
\newblock On quantum backpropagation, information reuse, and cheating measurement collapse, 2023.

\bibitem[Montanaro and Pallister(2016)]{Montanaro16}
Ashley Montanaro and Sam Pallister.
\newblock Quantum algorithms and the finite element method.
\newblock \emph{Phys. Rev. A}, 93:\penalty0 032324, Mar 2016.
\newblock \doi{10.1103/PhysRevA.93.032324}.
\newblock URL \url{https://link.aps.org/doi/10.1103/PhysRevA.93.032324}.

\bibitem[Ali and Kabel(2023)]{Ali23}
Mazen Ali and Matthias Kabel.
\newblock Performance study of variational quantum algorithms for solving the poisson equation on a quantum computer.
\newblock \emph{Phys. Rev. Appl.}, 20:\penalty0 014054, Jul 2023.
\newblock \doi{10.1103/PhysRevApplied.20.014054}.
\newblock URL \url{https://link.aps.org/doi/10.1103/PhysRevApplied.20.014054}.

\bibitem[Liu et~al.(2021)Liu, Arunachalam, and Temme]{Liu21}
Yunchao Liu, Srinivasan Arunachalam, and Kristan Temme.
\newblock A rigorous and robust quantum speed-up in supervised machine learning.
\newblock \emph{Nature Physics}, 17\penalty0 (9):\penalty0 1013--1017, September 2021.
\newblock ISSN 1745-2481.
\newblock \doi{10.1038/s41567-021-01287-z}.
\newblock URL \url{https://doi.org/10.1038/s41567-021-01287-z}.

\bibitem[Muser et~al.(2023)Muser, Zapusek, Belis, and Reiter]{Muser23}
Till Muser, Elias Zapusek, Vasilis Belis, and Florentin Reiter.
\newblock Provable advantages of kernel-based quantum learners and quantum preprocessing based on grover's algorithm, 2023.

\bibitem[Yamasaki et~al.(2023)Yamasaki, Isogai, and Murao]{Yamasaki23}
Hayata Yamasaki, Natsuto Isogai, and Mio Murao.
\newblock Advantage of quantum machine learning from general computational advantages, 2023.

\bibitem[Schuld and Killoran(2022)]{Schuld22}
Maria Schuld and Nathan Killoran.
\newblock Is quantum advantage the right goal for quantum machine learning?
\newblock \emph{PRX Quantum}, 3:\penalty0 030101, Jul 2022.
\newblock \doi{10.1103/PRXQuantum.3.030101}.
\newblock URL \url{https://link.aps.org/doi/10.1103/PRXQuantum.3.030101}.

\bibitem[Schuld et~al.(2021)Schuld, Sweke, and Meyer]{Schuld21}
Maria Schuld, Ryan Sweke, and Johannes~Jakob Meyer.
\newblock Effect of data encoding on the expressive power of variational quantum-machine-learning models.
\newblock \emph{Phys. Rev. A}, 103:\penalty0 032430, Mar 2021.
\newblock \doi{10.1103/PhysRevA.103.032430}.
\newblock URL \url{https://link.aps.org/doi/10.1103/PhysRevA.103.032430}.

\bibitem[Kordzanganeh et~al.(2023)Kordzanganeh, Sekatski, Fedichkin, and Melnikov]{Kordzanganeh23}
Mo~Kordzanganeh, Pavel Sekatski, Leonid Fedichkin, and Alexey Melnikov.
\newblock An exponentially-growing family of universal quantum circuits.
\newblock \emph{Machine Learning: Science and Technology}, 4\penalty0 (3):\penalty0 035036, aug 2023.
\newblock \doi{10.1088/2632-2153/ace757}.
\newblock URL \url{https://dx.doi.org/10.1088/2632-2153/ace757}.

\bibitem[Shin et~al.(2023)Shin, Teo, and Jeong]{Shin23}
S.~Shin, Y.~S. Teo, and H.~Jeong.
\newblock Exponential data encoding for quantum supervised learning.
\newblock \emph{Phys. Rev. A}, 107:\penalty0 012422, Jan 2023.
\newblock \doi{10.1103/PhysRevA.107.012422}.
\newblock URL \url{https://link.aps.org/doi/10.1103/PhysRevA.107.012422}.

\bibitem[K{\"u}bler* et~al.(2021)K{\"u}bler*, Buchholz*, and Sch{\"o}lkopf]{Kubler21}
J.~M. K{\"u}bler*, S.~Buchholz*, and B.~Sch{\"o}lkopf.
\newblock The inductive bias of quantum kernels.
\newblock In \emph{Advances in Neural Information Processing Systems 34 (NeurIPS 2021)}, pages 12661--12673. Curran Associates, Inc., December 2021.
\newblock URL \url{https://proceedings.neurips.cc/paper/2021/file/69adc1e107f7f7d035d7baf04342e1ca-Paper.pdf}.
\newblock *equal contribution.

\bibitem[Huang et~al.(2021{\natexlab{b}})Huang, Broughton, Mohseni, Babbush, Boixo, Neven, and McClean]{Huang21}
Hsin-Yuan Huang, Michael Broughton, Masoud Mohseni, Ryan Babbush, Sergio Boixo, Hartmut Neven, and Jarrod~R. McClean.
\newblock Power of data in quantum machine learning.
\newblock \emph{Nature Communications}, 12\penalty0 (1):\penalty0 2631, May 2021{\natexlab{b}}.
\newblock ISSN 2041-1723.
\newblock \doi{10.1038/s41467-021-22539-9}.
\newblock URL \url{https://doi.org/10.1038/s41467-021-22539-9}.

\bibitem[Jerbi et~al.(2023)Jerbi, Fiderer, Poulsen~Nautrup, Kübler, Briegel, and Dunjko]{Jerbi23}
Sofiene Jerbi, Lukas~J. Fiderer, Hendrik Poulsen~Nautrup, Jonas~M. Kübler, Hans~J. Briegel, and Vedran Dunjko.
\newblock Quantum machine learning beyond kernel methods.
\newblock \emph{Nature Communications}, 14\penalty0 (1):\penalty0 517, January 2023.
\newblock ISSN 2041-1723.
\newblock \doi{10.1038/s41467-023-36159-y}.
\newblock URL \url{https://doi.org/10.1038/s41467-023-36159-y}.

\bibitem[Cerezo et~al.(2023)Cerezo, Larocca, García-Martín, Diaz, Braccia, Fontana, Rudolph, Bermejo, Ijaz, Thanasilp, Anschuetz, and Holmes]{Cerezo23}
M.~Cerezo, Martin Larocca, Diego García-Martín, N.~L. Diaz, Paolo Braccia, Enrico Fontana, Manuel~S. Rudolph, Pablo Bermejo, Aroosa Ijaz, Supanut Thanasilp, Eric~R. Anschuetz, and Zoë Holmes.
\newblock Does provable absence of barren plateaus imply classical simulability? {O}r, why we need to rethink variational quantum computing, 2023.

\bibitem[Bowles et~al.(2023)Bowles, Wright, Farkas, Killoran, and Schuld]{Bowles23}
Joseph Bowles, Victoria~J Wright, Máté Farkas, Nathan Killoran, and Maria Schuld.
\newblock Contextuality and inductive bias in quantum machine learning, 2023.

\bibitem[Gili et~al.(2023)Gili, Alonso, and Schuld]{Gili23}
Kaitlin Gili, Guillermo Alonso, and Maria Schuld.
\newblock An inductive bias from quantum mechanics: learning order effects with non-commuting measurements, 2023.

\bibitem[Nguyen et~al.(2022)Nguyen, Schatzki, Braccia, Ragone, Coles, Sauvage, Larocca, and Cerezo]{Nguyen22}
Quynh~T. Nguyen, Louis Schatzki, Paolo Braccia, Michael Ragone, Patrick~J. Coles, Frederic Sauvage, Martin Larocca, and M.~Cerezo.
\newblock Theory for equivariant quantum neural networks, 2022.

\bibitem[Schatzki et~al.(2024)Schatzki, Larocca, Nguyen, Sauvage, and Cerezo]{Schatzki24}
Louis Schatzki, Martín Larocca, Quynh~T. Nguyen, Frédéric Sauvage, and M.~Cerezo.
\newblock Theoretical guarantees for permutation-equivariant quantum neural networks.
\newblock \emph{npj Quantum Information}, 10\penalty0 (1):\penalty0 12, January 2024.
\newblock ISSN 2056-6387.
\newblock \doi{10.1038/s41534-024-00804-1}.
\newblock URL \url{https://doi.org/10.1038/s41534-024-00804-1}.

\bibitem[Peters and Schuld(2023)]{Peters23}
Evan Peters and Maria Schuld.
\newblock Generalization despite overfitting in quantum machine learning models.
\newblock \emph{{Quantum}}, 7:\penalty0 1210, December 2023.
\newblock ISSN 2521-327X.
\newblock \doi{10.22331/q-2023-12-20-1210}.
\newblock URL \url{https://doi.org/10.22331/q-2023-12-20-1210}.

\bibitem[Cong et~al.(2019)Cong, Choi, and Lukin]{QCNN}
Iris Cong, Soonwon Choi, and Mikhail~D. Lukin.
\newblock Quantum convolutional neural networks.
\newblock \emph{Nature Physics}, 15\penalty0 (12):\penalty0 1273--1278, December 2019.
\newblock ISSN 1745-2481.
\newblock \doi{10.1038/s41567-019-0648-8}.
\newblock URL \url{https://doi.org/10.1038/s41567-019-0648-8}.

\bibitem[Meyer et~al.(2023)Meyer, Mularski, Gil-Fuster, Mele, Arzani, Wilms, and Eisert]{Meyer23}
Johannes~Jakob Meyer, Marian Mularski, Elies Gil-Fuster, Antonio~Anna Mele, Francesco Arzani, Alissa Wilms, and Jens Eisert.
\newblock Exploiting symmetry in variational quantum machine learning.
\newblock \emph{PRX Quantum}, 4:\penalty0 010328, Mar 2023.
\newblock \doi{10.1103/PRXQuantum.4.010328}.
\newblock URL \url{https://link.aps.org/doi/10.1103/PRXQuantum.4.010328}.

\bibitem[Umeano et~al.(2023)Umeano, Paine, Elfving, and Kyriienko]{Umeano23}
Chukwudubem Umeano, Annie~E. Paine, Vincent~E. Elfving, and Oleksandr Kyriienko.
\newblock What can we learn from quantum convolutional neural networks?, 2023.

\bibitem[Umeano et~al.(2024)Umeano, Elfving, and Kyriienko]{Umeano24}
Chukwudubem Umeano, Vincent~E. Elfving, and Oleksandr Kyriienko.
\newblock Geometric quantum machine learning of bqp$^a$ protocols and latent graph classifiers, 2024.

\bibitem[Kyriienko et~al.(2021)Kyriienko, Paine, and Elfving]{Kyriienko21}
Oleksandr Kyriienko, Annie~E. Paine, and Vincent~E. Elfving.
\newblock Solving nonlinear differential equations with differentiable quantum circuits.
\newblock \emph{Phys. Rev. A}, 103:\penalty0 052416, May 2021.
\newblock \doi{10.1103/PhysRevA.103.052416}.
\newblock URL \url{https://link.aps.org/doi/10.1103/PhysRevA.103.052416}.

\bibitem[Williams et~al.(2023)Williams, Paine, Wu, Elfving, and Kyriienko]{Williams23}
Chelsea~A. Williams, Annie~E. Paine, Hsin-Yu Wu, Vincent~E. Elfving, and Oleksandr Kyriienko.
\newblock Quantum chebyshev transform: Mapping, embedding, learning and sampling distributions, 2023.

\bibitem[Oseledets(2013)]{Oseledets13}
I.~V. Oseledets.
\newblock Constructive {Representation} of {Functions} in {Low}-{Rank} {Tensor} {Formats}.
\newblock \emph{Constructive Approximation}, 37\penalty0 (1):\penalty0 1--18, February 2013.
\newblock ISSN 1432-0940.
\newblock \doi{10.1007/s00365-012-9175-x}.
\newblock URL \url{https://doi.org/10.1007/s00365-012-9175-x}.

\bibitem[Ali and Nouy(2023)]{AliNouy23}
Mazen Ali and Anthony Nouy.
\newblock Approximation {Theory} of {Tree} {Tensor} {Networks}: {Tensorized} {Univariate} {Functions}.
\newblock \emph{Constructive Approximation}, 58\penalty0 (2):\penalty0 463--544, October 2023.
\newblock ISSN 1432-0940.
\newblock \doi{10.1007/s00365-023-09620-w}.
\newblock URL \url{https://doi.org/10.1007/s00365-023-09620-w}.

\bibitem[Braess(2007)]{Braess07}
Dietrich Braess.
\newblock \emph{Finite Elements: Theory, Fast Solvers, and Applications in Solid Mechanics}.
\newblock Cambridge University Press, 3 edition, 2007.

\bibitem[Yarotsky(2017)]{Yarostsky17}
Dmitry Yarotsky.
\newblock Error bounds for approximations with deep relu networks.
\newblock \emph{Neural Networks}, 94:\penalty0 103--114, 2017.
\newblock ISSN 0893-6080.
\newblock \doi{https://doi.org/10.1016/j.neunet.2017.07.002}.
\newblock URL \url{https://www.sciencedirect.com/science/article/pii/S0893608017301545}.

\bibitem[Nouy et~al.(2023)Nouy, Schueller, Jacquenot, MLasserre, and LEBRUN]{Nouy23}
Anthony Nouy, Julien Schueller, Guillaume Jacquenot, MLasserre, and Alice LEBRUN.
\newblock anthony-nouy/tensap: v1.5, July 2023.
\newblock URL \url{https://doi.org/10.5281/zenodo.8130836}.

\bibitem[Biamonte and Bergholm(2017)]{Biamonte17}
Jacob Biamonte and Ville Bergholm.
\newblock Tensor networks in a nutshell, 2017.

\bibitem[Schollwöck(2011)]{Schollwock11}
Ulrich Schollwöck.
\newblock The density-matrix renormalization group in the age of matrix product states.
\newblock \emph{Annals of Physics}, 326\penalty0 (1):\penalty0 96--192, 2011.
\newblock ISSN 0003-4916.
\newblock \doi{https://doi.org/10.1016/j.aop.2010.09.012}.
\newblock URL \url{https://www.sciencedirect.com/science/article/pii/S0003491610001752}.
\newblock January 2011 Special Issue.

\bibitem[Oseledets(2011)]{TT}
I.~V. Oseledets.
\newblock Tensor-train decomposition.
\newblock \emph{SIAM Journal on Scientific Computing}, 33\penalty0 (5):\penalty0 2295--2317, 2011.
\newblock \doi{10.1137/090752286}.
\newblock URL \url{https://doi.org/10.1137/090752286}.

\bibitem[Havlíček et~al.(2019)Havlíček, Córcoles, Temme, Harrow, Kandala, Chow, and Gambetta]{Havlicek19}
Vojtěch Havlíček, Antonio~D. Córcoles, Kristan Temme, Aram~W. Harrow, Abhinav Kandala, Jerry~M. Chow, and Jay~M. Gambetta.
\newblock Supervised learning with quantum-enhanced feature spaces.
\newblock \emph{Nature}, 567\penalty0 (7747):\penalty0 209--212, March 2019.
\newblock ISSN 1476-4687.
\newblock \doi{10.1038/s41586-019-0980-2}.
\newblock URL \url{https://doi.org/10.1038/s41586-019-0980-2}.

\bibitem[Ran(2020)]{Ran20}
Shi-Ju Ran.
\newblock Encoding of matrix product states into quantum circuits of one- and two-qubit gates.
\newblock \emph{Phys. Rev. A}, 101:\penalty0 032310, Mar 2020.
\newblock \doi{10.1103/PhysRevA.101.032310}.
\newblock URL \url{https://link.aps.org/doi/10.1103/PhysRevA.101.032310}.

\bibitem[Lubasch et~al.(2020)Lubasch, Joo, Moinier, Kiffner, and Jaksch]{Lubasch20}
Michael Lubasch, Jaewoo Joo, Pierre Moinier, Martin Kiffner, and Dieter Jaksch.
\newblock Variational quantum algorithms for nonlinear problems.
\newblock \emph{Phys. Rev. A}, 101:\penalty0 010301, Jan 2020.
\newblock \doi{10.1103/PhysRevA.101.010301}.
\newblock URL \url{https://link.aps.org/doi/10.1103/PhysRevA.101.010301}.

\bibitem[Gleinig and Hoefler(2021)]{Gleinig21}
Niels Gleinig and Torsten Hoefler.
\newblock An efficient algorithm for sparse quantum state preparation.
\newblock In \emph{2021 58th ACM/IEEE Design Automation Conference (DAC)}, pages 433--438, 2021.
\newblock \doi{10.1109/DAC18074.2021.9586240}.

\bibitem[Kazeev et~al.(2017)Kazeev, Oseledets, Rakhuba, and Schwab]{Kazeev17}
Vladimir Kazeev, Ivan Oseledets, Maxim Rakhuba, and Christoph Schwab.
\newblock {QTT}-finite-element approximation for multiscale problems {I}: model problems in one dimension.
\newblock \emph{Advances in Computational Mathematics}, 43\penalty0 (2):\penalty0 411--442, April 2017.
\newblock ISSN 1572-9044.
\newblock \doi{10.1007/s10444-016-9491-y}.
\newblock URL \url{https://doi.org/10.1007/s10444-016-9491-y}.

\bibitem[Bachmayr and Kazeev(2020)]{Bachmayr20}
Markus Bachmayr and Vladimir Kazeev.
\newblock Stability of {Low}-{Rank} {Tensor} {Representations} and {Structured} {Multilevel} {Preconditioning} for {Elliptic} {PDEs}.
\newblock \emph{Foundations of Computational Mathematics}, 20\penalty0 (5):\penalty0 1175--1236, October 2020.
\newblock ISSN 1615-3383.
\newblock \doi{10.1007/s10208-020-09446-z}.
\newblock URL \url{https://doi.org/10.1007/s10208-020-09446-z}.

\bibitem[Haberstich et~al.(2023)Haberstich, Nouy, and Perrin]{Haberstich23}
C\'{e}cile Haberstich, A.~Nouy, and G.~Perrin.
\newblock Active learning of tree tensor networks using optimal least squares.
\newblock \emph{SIAM/ASA Journal on Uncertainty Quantification}, 11\penalty0 (3):\penalty0 848--876, 2023.
\newblock \doi{10.1137/21M1415911}.
\newblock URL \url{https://doi.org/10.1137/21M1415911}.

\bibitem[Ritter et~al.(2024)Ritter, N\'u\~nez Fern\'andez, Wallerberger, von Delft, Shinaoka, and Waintal]{Ritter24}
Marc~K. Ritter, Yuriel N\'u\~nez Fern\'andez, Markus Wallerberger, Jan von Delft, Hiroshi Shinaoka, and Xavier Waintal.
\newblock Quantics tensor cross interpolation for high-resolution parsimonious representations of multivariate functions.
\newblock \emph{Phys. Rev. Lett.}, 132:\penalty0 056501, Jan 2024.
\newblock \doi{10.1103/PhysRevLett.132.056501}.
\newblock URL \url{https://link.aps.org/doi/10.1103/PhysRevLett.132.056501}.

\bibitem[Holmes and Matsuura(2020)]{Holmes20}
A.~Holmes and A.~Y. Matsuura.
\newblock Efficient quantum circuits for accurate state preparation of smooth, differentiable functions.
\newblock In \emph{2020 IEEE International Conference on Quantum Computing and Engineering (QCE)}, pages 169--179, Los Alamitos, CA, USA, oct 2020. IEEE Computer Society.
\newblock \doi{10.1109/QCE49297.2020.00030}.
\newblock URL \url{https://doi.ieeecomputersociety.org/10.1109/QCE49297.2020.00030}.

\bibitem[Melnikov et~al.(2023)Melnikov, Termanova, Dolgov, Neukart, and Perelshtein]{Melnikov23}
Ar~A Melnikov, A~A Termanova, S~V Dolgov, F~Neukart, and M~R Perelshtein.
\newblock Quantum state preparation using tensor networks.
\newblock \emph{Quantum Science and Technology}, 8\penalty0 (3):\penalty0 035027, jun 2023.
\newblock \doi{10.1088/2058-9565/acd9e7}.
\newblock URL \url{https://dx.doi.org/10.1088/2058-9565/acd9e7}.

\bibitem[Gourianov et~al.(2022)Gourianov, Lubasch, Dolgov, van~den Berg, Babaee, Givi, Kiffner, and Jaksch]{Gourianov22}
Nikita Gourianov, Michael Lubasch, Sergey Dolgov, Quincy~Y. van~den Berg, Hessam Babaee, Peyman Givi, Martin Kiffner, and Dieter Jaksch.
\newblock A quantum-inspired approach to exploit turbulence structures.
\newblock \emph{Nature Computational Science}, 2\penalty0 (1):\penalty0 30--37, January 2022.
\newblock ISSN 2662-8457.
\newblock \doi{10.1038/s43588-021-00181-1}.
\newblock URL \url{https://doi.org/10.1038/s43588-021-00181-1}.

\bibitem[Miyamoto and Ueda(2023)]{Miyamoto23}
Koichi Miyamoto and Hiroshi Ueda.
\newblock Extracting a function encoded in amplitudes of a quantum state by tensor network and orthogonal function expansion.
\newblock \emph{Quantum Information Processing}, 22\penalty0 (6):\penalty0 239, June 2023.
\newblock ISSN 1573-1332.
\newblock \doi{10.1007/s11128-023-03937-y}.
\newblock URL \url{https://doi.org/10.1007/s11128-023-03937-y}.

\bibitem[Iaconis et~al.(2024)Iaconis, Johri, and Zhu]{Iaconis24}
Jason Iaconis, Sonika Johri, and Elton~Yechao Zhu.
\newblock Quantum state preparation of normal distributions using matrix product states.
\newblock \emph{npj Quantum Information}, 10\penalty0 (1):\penalty0 15, January 2024.
\newblock ISSN 2056-6387.
\newblock \doi{10.1038/s41534-024-00805-0}.
\newblock URL \url{https://doi.org/10.1038/s41534-024-00805-0}.

\bibitem[Markeeva et~al.(2021)Markeeva, Tsybulin, and Oseledets]{markeeva2021}
L.~Markeeva, I.~Tsybulin, and I.~Oseledets.
\newblock Qtt-isogeometric solver in two dimensions.
\newblock \emph{Journal of Computational Physics}, 424:\penalty0 109835, 2021.
\newblock ISSN 0021-9991.
\newblock \doi{https://doi.org/10.1016/j.jcp.2020.109835}.
\newblock URL \url{https://www.sciencedirect.com/science/article/pii/S0021999120306094}.

\bibitem[Harrow et~al.(2009)Harrow, Hassidim, and Lloyd]{HHL1}
Aram~W. Harrow, Avinatan Hassidim, and Seth Lloyd.
\newblock Quantum algorithm for linear systems of equations.
\newblock \emph{Phys. Rev. Lett.}, 103:\penalty0 150502, Oct 2009.
\newblock \doi{10.1103/PhysRevLett.103.150502}.
\newblock URL \url{https://link.aps.org/doi/10.1103/PhysRevLett.103.150502}.

\bibitem[Childs et~al.(2017)Childs, Kothari, and Somma]{HHL2}
Andrew~M. Childs, Robin Kothari, and Rolando~D. Somma.
\newblock Quantum algorithm for systems of linear equations with exponentially improved dependence on precision.
\newblock \emph{SIAM Journal on Computing}, 46\penalty0 (6):\penalty0 1920--1950, 2017.
\newblock \doi{10.1137/16M1087072}.
\newblock URL \url{https://doi.org/10.1137/16M1087072}.

\bibitem[Bravo-Prieto et~al.(2023)Bravo-Prieto, LaRose, Cerezo, Subasi, Cincio, and Coles]{Bravo23}
Carlos Bravo-Prieto, Ryan LaRose, M.~Cerezo, Yigit Subasi, Lukasz Cincio, and Patrick~J. Coles.
\newblock Variational {Q}uantum {L}inear {S}olver.
\newblock \emph{{Quantum}}, 7:\penalty0 1188, November 2023.
\newblock ISSN 2521-327X.
\newblock \doi{10.22331/q-2023-11-22-1188}.
\newblock URL \url{https://doi.org/10.22331/q-2023-11-22-1188}.

\bibitem[Schuld and Petruccione(2021)]{schuld2021}
Maria Schuld and Francesco Petruccione.
\newblock \emph{Quantum Models as Kernel Methods}, pages 217--245.
\newblock Springer International Publishing, Cham, 2021.
\newblock ISBN 978-3-030-83098-4.
\newblock \doi{10.1007/978-3-030-83098-4_6}.
\newblock URL \url{https://doi.org/10.1007/978-3-030-83098-4_6}.

\end{thebibliography}

\end{document}